\documentclass[twocolumn, linenum]{aastex62}

\usepackage{CJK}
\usepackage{amsmath}
\CJKspace

\newcommand{\LCO}{\affiliation{Las Cumbres Observatory, 6740 Cortona Drive, Suite 102, Goleta, CA 93117-5575, USA}}
\newcommand{\UCSB}{\affiliation{Department of Physics, University of California, Santa Barbara, CA 93106-9530, USA}}

\newcommand{\UCD}{\affiliation{Department of Physics and Astronomy, University of California, Davis, 1 Shields Avenue, Davis, CA 95616-5270, USA}}

\newcommand{\STScI}{\affiliation{Space Telescope Science Institute, 3700 San Martin Drive, Baltimore, MD 21218-2410, USA}}

\newcommand{\IPAC}{\affiliation{Caltech/IPAC, Mailcode 100-22, Pasadena, CA 91125, USA}}

\newcommand{\UA}{\affiliation{Steward Observatory, University of Arizona, 933 North Cherry Avenue, Tucson, AZ 85721-0065, USA}}

\newcommand{\UNC}{\affiliation{Department of Physics and Astronomy, University of North Carolina, 120 East Cameron Avenue, Chapel Hill, NC 27599, USA}}

\newcommand{\JHU}{\affiliation{Department of Physics and Astronomy, The Johns Hopkins University, 3400 North Charles Street, Baltimore, MD 21218, USA}}

\newcommand{\GeminiNorth}{\affiliation{Gemini Observatory, 670 North A`ohoku Place, Hilo, HI 96720-2700, USA}}
\newcommand{\Keck}{\affiliation{W.~M.~Keck Observatory, 65-1120 M\=amalahoa Highway, Kamuela, HI 96743-8431, USA}}
\newcommand{\UW}{\affiliation{Department of Astronomy, University of Washington, 3910 15th Avenue NE, Seattle, WA 98195-0002, USA}}
\newcommand{\Catalyst}{\altaffiliation{LSSTC Catalyst Fellow}}
\newcommand{\USask}{\affiliation{Department of Physics and Engineering Physics, University of Saskatchewan, 116 Science Place, Saskatoon, SK S7N 5E2, Canada}}

\newcommand{\Rutgers}{\affiliation{Department of Physics and Astronomy, Rutgers, the State University of New Jersey,\\136 Frelinghuysen Road, Piscataway, NJ 08854-8019, USA}}
\newcommand{\FSU}{\affiliation{Department of Physics, Florida State University, 77 Chieftan Way, Tallahassee, FL 32306-4350, USA}}

\newcommand{\TAMU}{\affiliation{Department of Physics and Astronomy, Texas A\&M University, 4242 TAMU, College Station, TX 77843, USA}}

\newcommand{\ICE}{\affiliation{Institute of Space Sciences (ICE, CSIC), Campus UAB, Carrer
de Can Magrans, s/n, E-08193 Barcelona, Spain}}
\newcommand{\IEEC}{\affiliation{Institut d'Estudis Espacials de Catalunya, Gran Capit\`a, 2-4, Edifici Nexus, Desp.\ 201, E-08034 Barcelona, Spain}}

\newcommand{\MSU}{\affiliation{Center for Data Intensive and Time Domain Astronomy, Department of Physics and Astronomy,\\Michigan State University, East Lansing, MI 48824, USA}}
\newcommand{\IAP}{\affiliation{Institut d'Astrophysique de Paris, CNRS-Sorbonne Universit\'e, 98 bis boulevard Arago, 75014 Paris, France}}
\newcommand{\Pitt}{\affiliation{Department of Physics and Astronomy \& Pittsburgh Particle Physics, Astrophysics, and Cosmology Center (PITT PACC), University of Pittsburgh, 3941 O'Hara Street, Pittsburgh, PA 15260, USA}}
\newcommand{\Vtech}{\affiliation{Department of Physics, Virginia Tech, Blacksburg, VA 24061, USA}}
\newcommand{\IAC}{\affiliation{Instituto de Astrof{\'\i}sica de Canarias, E-38205 La Laguna, Tenerife, Spain}}
\newcommand{\Laguna}{\affiliation{Universidad de La Laguna, Dept. Astrof{\'\i}sica, E-38206 La Laguna, Tenerife, Spain}}
\newcommand{\UOak}{\affiliation{Homer L. Dodge Department of Physics and Astronomy, University of Oklahoma, 440 W. Brooks, Rm 100, Norman, OK 73019-2061, USA}}
\newcommand{\Hamburg}{\affiliation{Hamburger Sternwarte, Gojenbergsweg 112, D-21029 Hamburg, Germany}}
\newcommand{\PSI}{\affiliation{Planetary Science Institute, 1700 East Fort Lowell, Suite 106, Tucson, AZ 85719-2395 USA}}
\newcommand{\SETI}{\affiliation{SETI Institute, 339 Bernardo Ave, Suite 200, Mountain View, CA 94043, USA}}

\newcommand{\msun}[1]{$\mathrm{M_{\odot}}$}
\newcommand{\rsun}[1]{$\mathrm{R_{\odot}}$}
\newcommand{\lsun}[1]{$\mathrm{L_{\odot}}$}
\newcommand{\Halpha}[1]{$\mathrm{H\alpha}$}
\newcommand{\funit}{\hbox{erg cm$^{-2}$ s$^{-1}$ \AA$^{-1}$}}

\defcitealias{astropy_collaboration_astropy_2013}{Astropy Collaboration 2013}
\defcitealias{astropy_collaboration_astropy_2018}{2018}
\defcitealias{astropy_collaboration_astropy_2022}{2022}
\defcitealias{center_heasarc_heasoft_2014}{Heasarc, 2014}

\graphicspath{{./}{figures/}}

\shorttitle{Type II FUV Spectrum}
\shortauthors{Bostroem et al.}

\begin{document}

\title{SN~2022acko: the First Early Far-Ultraviolet Spectra of a Type IIP Supernova}

\correspondingauthor{K. Azalee Bostroem}
\email{bostroem@arizona.edu}
\author[0000-0002-4924-444X]{K.\ Azalee Bostroem}
\Catalyst\UA
\author[0000-0003-0599-8407]{Luc Dessart}
\IAP
\author{D. John Hillier}
\Pitt
\author[0000-0001-9589-3793]{Michael Lundquist}
\Keck
\author[0000-0003-0123-0062]{Jennifer E.\ Andrews}
\GeminiNorth
\author[0000-0003-4102-380X]{David J.\ Sand}
\UA
\author[0000-0002-7937-6371]{Yize Dong \begin{CJK*}{UTF8}{gbsn}(董一泽)\end{CJK*}}
\UCD
\author[0000-0001-8818-0795]{Stefano Valenti}
\UCD
\author[0000-0002-6703-805X]{Joshua Haislip}
\UNC
\author[0000-0003-2744-4755]{Emily T. Hoang}
\UCD
\author[0000-0002-0832-2974]{Griffin Hosseinzadeh}
\UA
\author[0000-0003-0549-3281]{Daryl Janzen}
\USask
\author[0000-0001-5754-4007]{Jacob E.\ Jencson}
\JHU
\author[0000-0001-8738-6011]{Saurabh W.\ Jha}
\Rutgers
\author[0000-0003-3642-5484]{Vladimir Kouprianov}
\UNC
\author[0000-0002-0744-0047]{Jeniveve Pearson}
\UA
\author[0000-0002-7015-3446]{Nicolas E.\ Meza Retamal}
\UCD
\author[0000-0002-5060-3673]{Daniel E.\ Reichart}
\UNC
\author[0000-0002-4022-1874]{Manisha Shrestha}
\UA
\author[0000-0002-5221-7557]{Christopher Ashall}
\Vtech
\author[0000-0001-5393-1608]{E.~Baron}
\PSI
\Hamburg
\UOak
\author[0000-0001-6272-5507]{Peter J.\ Brown}
\TAMU
\author[0000-0002-7566-6080]{James~M.~DerKacy}
\Vtech
\author[0000-0003-4914-5625]{Joseph Farah}
\LCO\UCSB
\author[0000-0002-1296-6887]{Llu\'is Galbany}
\ICE
\IEEC
\author[0000-0002-0264-7356]{J. I. Gonz\'alez Hern\'andez}
\IAC
\Laguna
\author{Elizabeth Green}
\UA
\author[0000-0002-4338-6586]{Peter Hoeflich}
\FSU
\author[0000-0003-4253-656X]{D.\ Andrew Howell}
\LCO\UCSB
\author[0000-0003-3108-1328]{Lindsey A.\ Kwok}
\Rutgers
\author[0000-0001-5807-7893]{Curtis McCully}
\LCO\UCSB
\author[0000-0003-3939-7167]{Tom\'as E. M\"uller-Bravo}
\ICE
\IEEC
\author[0000-0001-9570-0584]{Megan Newsome}
\LCO\UCSB
\author[0000-0003-0209-9246]{Estefania Padilla Gonzalez}
\LCO\UCSB
\author[0000-0002-7472-1279]{Craig Pellegrino}
\LCO\UCSB
\author[0000-0003-3643-839X]{Jeonghee Rho}
\SETI
\author{Micalyn Rowe}
\TAMU
\author[0009-0002-5096-1689]{Michaela Schwab}
\Rutgers
\author[0000-0002-9301-5302]{Melissa Shahbandeh}
\STScI\JHU
\author[0000-0001-5510-2424]{Nathan Smith}
\UA
\author[0000-0002-1468-9668]{Jay Strader}
\MSU
\author[0000-0003-0794-5982]{Giacomo Terreran}
\LCO
\author[0000-0001-9038-9950]{Schuyler D. Van Dyk}
\IPAC
\author[0000-0003-2732-4956]{Samuel Wyatt}
\UW

\begin{abstract} 
We present five far- and near-ultraviolet spectra of the Type II plateau supernova, SN~2022acko, obtained 5, 6, 7, 19, and 21 days after explosion, all observed with the Hubble Space Telescope/Space Telescope Imaging Spectrograph.
The first three epochs are earlier than any Type II plateau supernova has been observed in the far-ultraviolet revealing unprecedented characteristics.
These three spectra are dominated by strong lines, primarily from metals, which contrasts with the featureless early optical spectra.
The flux decreases over the initial time series as the ejecta cool and line-blanketing takes effect.
We model this unique dataset with the non-local thermodynamic equilibrium radiation transport code \texttt{CMFGEN}, finding a good match to the explosion of a low-mass red supergiant with energy $E_{\mathrm{kin}}=6\times10^{50}\mathrm{\,erg}$.
With these models we identify, for the first time, the ions that dominate the early ultraviolet spectra.

We present optical photometry and spectroscopy, showing that SN~2022acko has a peak absolute magnitude of $V=-15.4$ mag and plateau length of $\sim115$d.
The spectra closely resemble those of SN~2005cs and SN~2012A.
Using the combined optical and ultraviolet spectra, we report the fraction of flux as a function of bluest wavelength on days 5, 7, and 19.
We create a spectral time-series of Type II supernovae in the ultraviolet, demonstrating the rapid decline of flux over the first few weeks of evolution.  
Future observations of Type II supernovae are required to map out the landscape of exploding red supergiants, with and without circumstellar material, which is best revealed in high-quality ultraviolet spectra. 
\end{abstract}


\section{Introduction} \label{sec:intro}
Hydrogen-rich supernovae, also known as Type II supernovae, are thought to come from stars with masses $\gtrsim$8~\msun{} \citep[e.g.][]{woosley_physics_1986}.
While massive stars and their supernova explosions play a fundamental role in the evolution of the universe, many details of their evolution are not understood. 
Observations at various wavelengths and over multiple epochs can probe different aspects of the supernova explosion \citep[e.g.][]{dessart_quantitative_2005, fransson_physical_1984, jencson_SPIRITS_2019}. 
While complete optical coverage is becoming more common, ultraviolet (UV) spectroscopy remains almost non-existent. 

Although the UV is not well studied, these wavelengths contain a wealth of information.
In contrast to the nearly featureless early optical, the early UV, and in particular the far-UV (FUV; $\lambda <1700$ \AA{}) is full of metal features.
These features can be used to determine the composition of the outer stellar envelope or circumstellar material (CSM; material lost just prior to explosion), the density and temperature of the outer layers of the ejecta, as well as interaction between any CSM and the supernova photons and/or ejecta \citep[e.g.][]{panagia_coordinated_1980, fransson_physical_1984, fransson_implications_1987, dessart_quantitative_2005, dessart_modeling_2022}.
Additionally, as current and future missions such as JWST and the Nancy Grace Roman Space Telescope focus on the infrared (IR), understanding the UV spectra of massive star supernovae is key for interpreting the early universe where UV photons from the first supernovae are redshifted into the IR.

This lack of UV spectroscopy is due to two competing challenges.
First, UV emission in Type II supernovae is powered by the energy deposited in the hydrogen-rich envelope by the shock which fades rapidly as it cools \citep{pritchard_bolometric_2014}.
This effect is enhanced by the large number of iron absorption features which appear in the first few weeks after explosion and blend together to absorb most of the UV flux \citep{pun_ultraviolet_1995, gezari_probing_2008, dessart_using_2008, dessart_supernova_2010}.
Although all UV spectra eventually show this evolution, the timing and speed of the damping appear to vary (see \autoref{subsec:comparison}).
Second, the UV can only be observed from space and the instruments best suited for these observations, the Space Telescope Imaging Spectrograph (STIS) and the Cosmic Origins Spectrograph (COS), are on the Hubble Space Telescope (HST), whose scheduling is not designed for rapid target of opportunity (ToO) observations.
Nevertheless, a limited number of disruptive ToOs are designated each year with a turn around time of 2-5 days.
Adding an additional barrier to early observations is the fact that every HST detector capable of UV observations, other than the STIS/CCD (which only covers NUV wavelengths), must pass a bright object screening so as not to damage the detector.

The only FUV observations of Type II supernovae taken earlier than the spectra presented in this paper are from the International Ultraviolet Explorer (IUE) and are of atypical Type II supernovae (1980K; \citealt{pettini_interstellar_1982},  1987A; \citealt{pun_ultraviolet_1995, cassatella_spectral_1987, fransson_implications_1987}).
The earliest FUV spectrum of a non-interacting normal Type II supernova is SN~1999em, observed $\sim$12 days after explosion \citep{baron_preliminary_2000}.
In the NUV there are a handful of early spectroscopic observations (SN~2005cs: \citealt[][]{brown_early_2007}, SN~2021yja: \citealt{vasylyev_early-time_2022}, SN~2020fqv: \citealt{tinyanont_progenitor_2022}, SN~2022wsp: \citealt{vasylyev_early-time_2023}, SN~2005ay: \citealt{gal-yam_galex_2008}, SN~2013ej: \citealt{dhungana_extensive_2016}), however these spectra are either low S/N, observed over a week post explosion, or suffer from significant reddening. 
While \citet{gal-yam_galex_2008} noted the similarity of the NUV features of supernovae observed with GALEX, despite the optical diversity, the sample is based on three supernovae taken more than a week after explosion and did not extend to the FUV. 
The putative uniformity of the NUV spectra of Type II supernovae differs from the diversity observed in stripped-envelope supernovae \citep{kwok_ultraviolet_2022}.
A larger sample with better time sampling and signal-to-noise (S/N) is required to assess the true diversity of this wavelength range. 

With the limited lifetime of HST and the dearth of UV observations, we proposed in Cycle 30 to gather early FUV spectra of a Type II supernova (PI: Bostroem; GO-17132). 
We limited our selection to the most common kind of Type II supernovae, those that display either a plateau (IIP-like) or linear decline (IIL-like) in their optical light curves after peak and a rapid drop onto the radioactive tail, $\sim$80-120 days after explosion \citep{Valenti2016}.

On 2022 December 6 (JD 2459919.59), the Distance Less than 40 Mpc (DLT40) survey \citep{tartaglia_early_2018}, discovered SN~2022acko in NGC~1300 \citep{lundquist_dlt40_2022}.
A day later we activated a disruptive ToO with HST to obtain STIS FUV and NUV spectra in a three-day time series, with a final epoch about two weeks later.
Our observations executed within three days of triggering, resulting in the first FUV observations of a Type IIP supernova within a week of explosion (see \autoref{sec:discovery} for definition of explosion epoch), significantly earlier than the spectrum of SN~1999em taken $\sim$12 days after explosion.

In this paper we present our UV dataset, using optical data to place SN~2022acko in the context of other Type II supernovae.
We describe the HST triggering process in \autoref{sec:discovery}.
In \autoref{sec:obs} we present photometric and spectroscopic observations. 
The optical light curve and spectroscopic evolution are described in \autoref{sec:analysis}.
Finally, we describe the UV spectra and modeling in \autoref{sec:uv} before presenting our conclusions in \autoref{sec:conclusion}.

\section{HST Trigger and Scheduling} \label{sec:discovery}
\begin{figure*}
    \centering
    \includegraphics{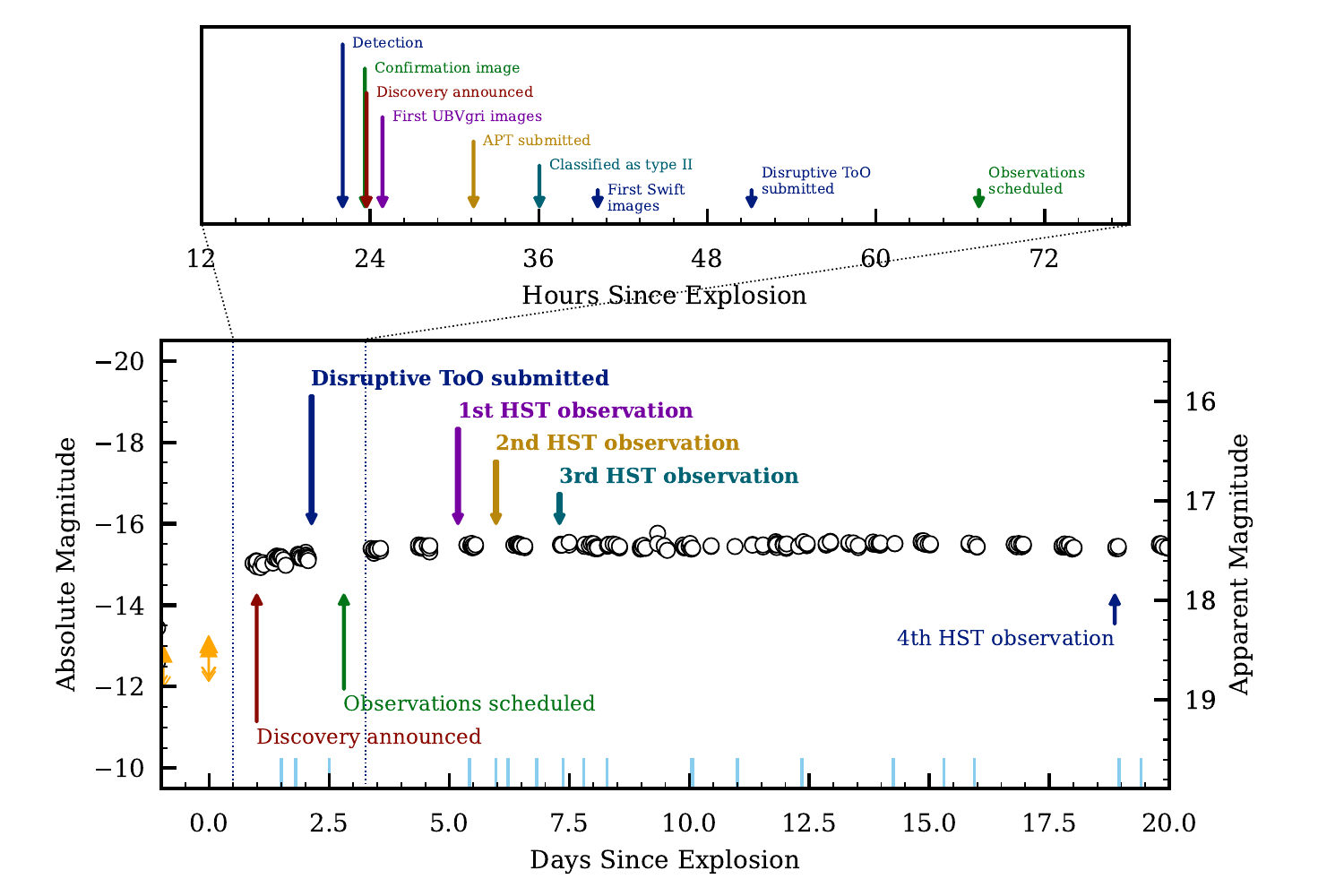}
    \caption{A timeline of events from discovery through the end of HST observations for SN~2022acko. 
    The \textit{bottom} panel shows the DLT40 open filter light curve (open circles) and the ATLAS limits (orange triangles).
    Key moments in the timeline are marked on the light curve. 
    The epochs at which optical spectra were obtained are shown as light blue lines at the bottom of the figure. 
    We observed at least one optical spectrum concurrent with each HST visit.
    The \textit{top} panel shows the details of activities executed in the first three days after explosion to enable the triggering of HST. 
    If any one of these events was unable to occur or was delayed, we would not have been able to observe SN~2022acko in the FUV.}
    \label{fig:timeline}
\end{figure*}

Early FUV spectral sequences are rare and have not been obtained for a Type IIP supernova because they require a nearby object, discovered and characterized early, with low extinction, in a field without other bright objects. 
In this section we describe the process of discovering, vetting, and triggering SN~2022acko for early HST FUV and NUV observations. \autoref{fig:timeline} summarizes this process, with the first three days shown in detail in the top panel, from detection through scheduling, and the full sequence in the bottom panel.

SN~2022acko was discovered by the DLT40 team \citep{tartaglia_early_2018} on UT 2022-12-06 03:53:00 \newline (JD 2459919.66; r=16.54 mag). 
Within 15 minutes of discovery, it was confirmed with a second image, multi-band imaging and spectroscopy were triggered, and the supernova was announced to the community via the Transient Name Server\footnote{\url{https://www.wis-tns.org}} \citep[TNS;][]{lundquist_dlt40_2022}.
Using the automated infrastructure of the DLT40 survey \citep{yang_optical_2019},  we immediately identified a deep ATLAS non-detection within 24 hours of discovery (JD 2459918.67, $o>19.1$ mag; \citealt{tonry_atlas_2018, smith_prime_2006}).
Conservatively, we use this non-detection as the explosion epoch for the remainder of the paper.
The combination of the ATLAS non-detection and DLT40 discovery within 24 hours allowed us to trigger our Neil Gehrels Swift Observatory \citep{gehrels_swift_2004} GI proposal (PI: Dong; 1821204) to obtain six-hour cadence UV and optical images of the supernova.

We received our first Swift photometry 14.5 hours after discovery, verifying that the supernova was UV bright and vetting the field for UV bright sources using the \textit{uvw2} NUV filter image.
SN~2022acko was classified as a Type II supernova 24 hours after discovery \citep{li_lions_2022}, which was subsequently verified an hour later with another spectrum \citep{meza_dlt40_2022}.
We submitted the official HST disruptive ToO trigger 28 hours after the discovery of SN~2022acko.
The bright object protection (BOP) review was completed 44 hours after discovery, and the ToO was scheduled to execute $\sim2.5$ days later.

The first three epochs were executed 5.2, 6.0, and 7.3 days after explosion (JD 2459923.87, 2459924.66, and 2459925.98), respectively. 
These constitute the earliest FUV observations of a Type IIP/L supernova to date and some of the earliest NUV observations.
Scheduling constraints prevented our final epoch from being taken a week after the initial sequence and we thus observed SN~2022acko 18.9 days after explosion (JD 2459937.53), completing our proposed sequence.

Unfortunately, the second epoch failed to acquire guide stars and only one FUV exposure was taken. 
We immediately notified the Space Telescope Science Institute (STScI) of the failure and were awarded an additional orbit.
Due to the same scheduling constraints that limited the execution of the fourth epoch, the make-up epoch executed 20.7 days after explosion (JD 2459938.89).
Given the expected low FUV flux, we devoted this epoch exclusively to the NUV.

\section{Observations}\label{sec:obs}
We triggered high-cadence photometric and spectroscopic follow-up observations immediately following the discovery of SN~2022acko with both ground and space based instruments. Below we report on the facilities used and the data that were collected over this campaign. 

\subsection{Photometry}
We obtained \textit{UBgVri}-band photometry from the Las Cumbres Observatory's network of 0.4m, 1m, and 2m telescopes \citep{brown_cumbres_2013} through the Global Supernova Project (GSP) collaboration and \textit{BgVri}-band and \textit{Open} filter photometry from SkyNet's network of 0.4m PROMPT telescopes \citep{reichart_prompt_2005} through the DLT40 collaboration. 
The multi-band images from SkyNet are a new capability which enable the DLT40 collaboration to obtain very high-cadence coverage over the optical wavelength range.
This is the first published light curve to utilize this new feature.

Las Cumbres Observatory images were bias corrected and flat fielded using BANZAI \citep{mccully_real-time_2018} then PSF-photometry was performed in an IRAF-based pipeline \citep{Valenti2016}. 
The SkyNet images were processed and aperture-photometry performed in a python-based pipeline for all color filters. 
\textit{UBV} magnitudes are presented in Vega magnitudes while \textit{gri}-band magnitudes are given in AB magnitudes.
The \textit{Open} filter SkyNet images were reduced with difference imaging using \texttt{HOTPANTS} \citep{becker_hotpants_2015} prior to magnitudes being extracted and then calibrated to \textit{r}-band.
We supplement these observations with photometry downloaded from the ATLAS forced photometry service in the \textit{orange} filter \citep{smith_prime_2006, tonry_atlas_2018}.

UV and optical imaging was also obtained with the UVOT instrument \citep{roming_swift_2005} on Swift in the 
\textit{uvw2}, \textit{uvm2}, \textit{uvw1}, \textit{u}, \textit{b}, and \textit{v} filters. 
The rapid response of Swift provided key information on the UV brightness of SN~2022acko, essential for triggering HST, and caught the rarely seen NUV light-curve rise. 
Images were reduced with aperture photometry using the High-energy Astrophysics Software \texttt{UVOT} routines \citepalias{center_heasarc_heasoft_2014} with the updated sensitivity of \citet{breeveld_updated_2011}. 
We used a 5.4\arcsec{} aperture set to encompass the full light of the supernova and a region near the supernova was selected and used to subtract the background.
Photometry is automatically corrected for aperture and coincidence losses.
The final UV images are upper limits only, as the supernova has faded below the background flux of the host galaxy. 
It is possible that more detections can be recovered with difference imaging when late-time templates are obtained.
The late-time UV photometry is sensitive to contamination by the underlying host galaxy and therefore both the object and background aperture must be carefully selected.
Experimenting with different aperture sizes and background locations, we find that this primarily affects the Swift UV limits.

\subsection{Spectroscopy}
We obtained ground-based optical spectra with the FLOYDS spectrograph on the Las Cumbres Observatory 2m Faulkes Telescopes North and South (FTN \& FTS; \citealt{brown_cumbres_2013}) through the GSP collaboration.

\begin{figure*}
    \centering
    \includegraphics{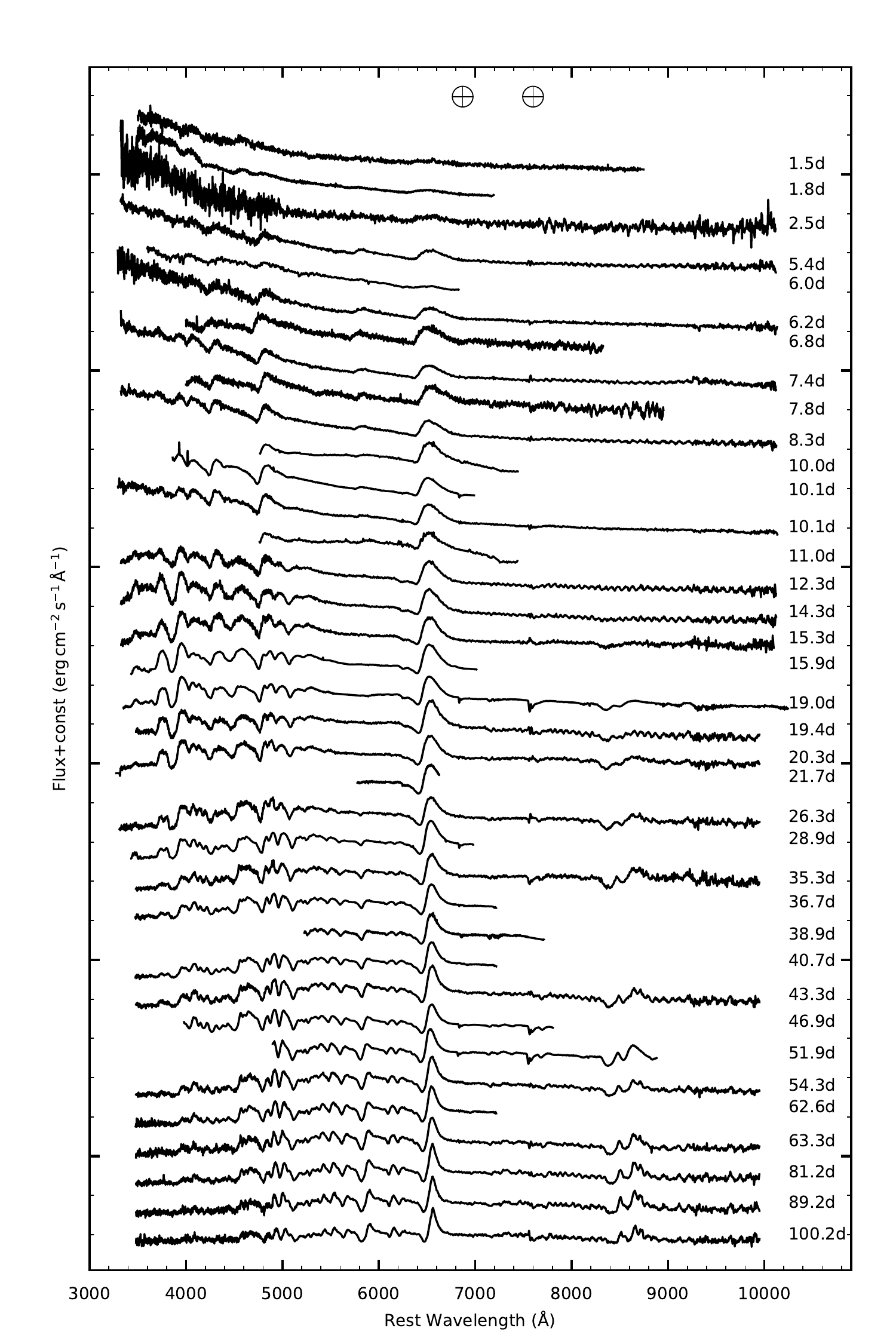}
    \caption{The spectroscopic evolution of SN~2022acko in the optical.
    The phase of each spectrum, relative to the date of last non-detection, is marked on the right.
    Telluric features are marked at the top of the figure.
    The spectra shown in this figure are available as the Data behind the Figure.}
    \label{fig:OptSpecMontage}
\end{figure*}

Additional spectra were obtained with the Binospec spectrograph on the MMT \citep{fabricant_binospec_2019}, the Boller and Chivens (B\&C) spectrograph\footnote{\url{http://james.as.arizona.edu/~psmith/90inch/bcman/html/bcman.html}} on the Bok Telescope, 
the Robert Stobie Spectrograph (RSS) on the Southern African Large Telescope (SALT; \citealt{smith_prime_2006}), the Goodman Spectrograph on the Southern Astrophysical Research Telescope (SOAR; \citealt{clemens_goodman_2004}), Multi-Object Double Spectrograph (MODS; \citealt{pogge_multi-object_2010}) on the Large Binocular Telescope (LBT; \citealt{hill_large_2010}), the Intermediate Dispersion Spectrograph (IDS) at the Isaac Newton Telescope (INT).  Other spectra were downloaded directly from the TNS.
A complete list of spectroscopic observations is given in \autoref{tab:spec} and shown in \autoref{fig:OptSpecMontage}.
\begin{deluxetable}{cccr}
\tablecaption{Spectroscopic observations of SN~2022acko\label{tab:spec}}
\tablehead{\colhead{Observer-frame } & \colhead{JD} & \colhead{Telescope} & \colhead{Instrument}\\
\colhead{Phase (d)} & \colhead{} & \colhead{} & \colhead{}}
\startdata
    1.5   & 2459920.17 & Lijiang-2.4m & YFOSC      \\ 
    1.8   & 2459920.47 & SALT         & RSS       \\ 
    2.5   & 2459921.17 & FTS          & FLOYDS    \\ 
    5.2   & 2459923.87  & HST          & STIS      \\ 
    5.4   & 2459924.10 & FTS          & FLOYDS    \\ 
    6.0   & 2459924.65 & Bok          & B\&C        \\  
    6.0   & 2459924.66 & HST          & STIS       \\ 
    6.2   & 2459924.90 & FTN          & FLOYDS    \\ 
    6.8   & 2459925.50 & INT          & IDS      \\ 
    7.3   & 2459925.98 & HST          & STIS      \\ 
    7.4   & 2459926.04 & FTS          & FLOYDS    \\ 
    7.8   & 2459926.50 & INT          & IDS      \\ 
    8.3   & 2459926.96 & FTS          & FLOYDS     \\
    10.0  & 2459928.72 & Bok          & B\&C      \\ 
    10.1  & 2459928.73 & SOAR         & Goodman    \\ 
    10.1  & 2459928.75 & FTN          & FLOYDS     \\ 
    11.0  & 2459929.67 & Bok          & B\&C      \\ 
    12.3  & 2459931.02 & FTS          & FLOYDS     \\ 
    14.3  & 2459932.92 & FTS          & FLOYDS     \\ 
    15.3  & 2459933.98 & FTS          & FLOYDS     \\ 
    15.9  & 2459934.61 & SOAR         & Goodman   \\ 
    18.9  & 2459937.54 & HST          & STIS      \\
    19.0  &  2459937.63 & LBT         & MODS      \\ 
    19.4  & 2459938.08 & FTS          & FLOYDS     \\ 
    20.3  & 2459938.99 & FTS          & FLOYDS    \\ 
    20.7  & 2459939.39 & HST          & STIS      \\ 
    21.7   & 2459940.42 & SALT         & RSS       \\ 
    28.9  & 2459944.97 & FTS          & FLOYDS    \\ 
    35.3  & 2459947.62 & SOAR         & Goodman   \\ 
    35.2  & 2459953.94 & FTS          & FLOYDS    \\ 
    36.7  & 2459955.40 & SALT         & RSS       \\ 
    38.9  & 2459957.60 & MMT          & Binospec  \\ 
    40.7  & 2459959.38 & SALT         & RSS       \\ 
    43.3  & 2459961.98 & FTS          & FLOYDS    \\ 
    46.9  & 2459965.58 & SOAR         & Goodman     \\ 
    51.9  & 2459970.56 & SOAR         & Goodman    \\ 
    54.3  & 2459972.95 & FTS          & FLOYDS    \\ 
    62.6  & 2459981.31 & SALT         & RSS       \\ 
    63.3  & 2459981.95 & FTS          & FLOYDS    \\ 
    81.2  & 2459999.90 & FTS          & FLOYDS    \\ 
    89.2  & 2460007.90 & FTS          & FLOYDS    \\ 
    100.2 & 2460018.90 & FTS          & FLOYDS    \\ 
\enddata
\end{deluxetable}

Most spectroscopic observations were reduced with  \texttt{IRAF} (B\&C) or IRAF-based pipelines (FLOYDS: \citealt{valenti_first_2014}; MODS: \citealt{pogge_rwpoggemodsccdred_2019}) using standard reduction techniques.
RSS observations were reduced with a custom pipeline using the \texttt{PySALT} software package \citep{crawford_pysalt_2010}, Binospec was reduced with a custom IDL pipeline \citep{kansky_binospec_2019}, IDS spectra were reduced with the custom python package IDSRED \citep{tomas_e_muller_bravo_2023_7851772}, and Goodman spectra were reduced with a custom python package \footnote{\url{https://soardocs.readthedocs.io/projects/goodman-pipeline/en/latest/}}. All spectra were scaled to the Swift photometry and \textit{gri}-band photometry using either a constant offset or linear fit.

\subsection{HST UV Spectroscopic Observations}
FUV and NUV spectra were obtained with HST/STIS 5.2, 6.0, 7.3, and 18.9 days after explosion. 
A final NUV spectrum was obtained with STIS on day 20.8.
Observations were taken with the G140L and G230L gratings using the FUV- and NUV-MAMA detectors, respectively, with the 52$\times$0.2\arcsec{} slit.
Exposure time, resolution, and wavelength range for these observations are given in \autoref{tab:HSTspec}. 
Reduced observations were downloaded from the Mikulski Archive for Space Telescopes (MAST). 
The data described here may be obtained from the MAST archive at
\dataset[doi:10.17909/gaze-k021]{https://dx.doi.org/10.17909/gaze-k021}.

\begin{deluxetable*}{ccccccr}
\tablecaption{HST/STIS observations of SN~2022acko\label{tab:HSTspec}}
\tablehead{\colhead{Observer-frame Phase (d)} & \colhead{JD} & \colhead{Grating } & \colhead{Exposure Time (s)} &  \colhead{Resolving Power} & \colhead{Wavelength Range (\AA)}}
\startdata
    5.2   & 2459923.9  &    G140L          & 3000 & 1000&1150--1730  \\ 
    5.3   & 2459924.0  &    G230L         & 1500 & 500 & 1570--3180 \\ 
    6.0   & 2459924.7 &    G140L          & 2250 & 1000 & 1150--1730   \\ 
    7.3   & 2459926.0 &    G140L          & 1465 & 1000 & 1150--1730  \\ 
    7.3   & 2459926.0 &    G230L          & 440 & 500 & 1570--3180 \\ 
    18.9  & 2459937.6 &    G140L          & 6604 & 1000 & 1150--1730 \\
    19.0  & 2459937.7 &    G230L          & 757 & 500 &  1570--3180\\
    20.8  & 2459939.4 &    G230L          & 2278 & 500 &  1570--3180\\ 
\enddata
\end{deluxetable*}

\section{Analysis}\label{sec:analysis}
\begin{figure*}
    \centering
    \includegraphics[width=\textwidth]{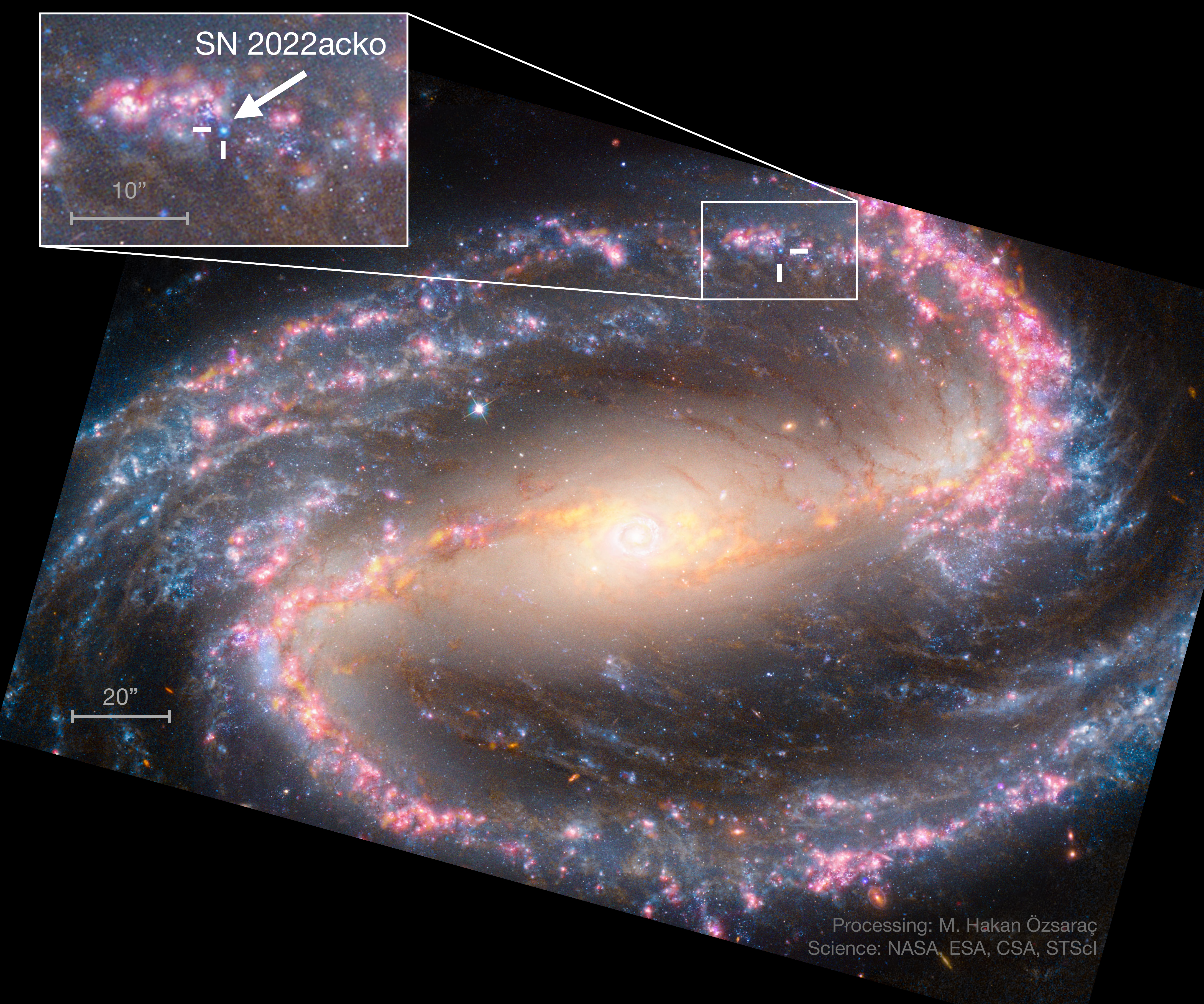}
    \caption{JWST (blue), HST (blue), VLT (red and magneta), and ALMA (yellow)} composite image of NGC~1300. 
    SN~2022acko lies in a star-forming region in a spiral arm north east of the nucleus and is marked in both the image and the inset with white ticks.
    Image adapted from M. Hakan \"{O}zsara\c{c} and Alyssa Pagan, NASA, ESA, ESO-Chile, ALMA, NAOJ, NRAO; north up, east right. 
    \label{fig:JWST}
\end{figure*}

SN~2022acko exploded in a spiral arm of NGC~1300 \citep[$z=0.00526$;][]{springob_digital_2005}, a well-studied nearby galaxy. 
\autoref{fig:JWST}\footnote{\url{https://www.flickr.com/photos/mhozsarac/52770584971/}} shows the location of SN~2022acko in a star-forming region of NGC~1300, in a composite JWST, HST, Very Large Telescope (VLT)/Multi-Unit Spectroscopic Explorer (MUSE), and Atacama Large Millimeter/submillimeter Array (ALMA) image.
For the analysis of SN~2022acko, we must first define the distance to the host galaxy as well as galactic and host extinction. 
These, along with other fundamental parameters, are summarized in \autoref{tab:prop}.

\begin{deluxetable}{lr}
\tablecaption{Properties of SN~2022acko\label{tab:prop}}
\tablehead{\colhead{Property} & \colhead{Value} }
\startdata
    RA & $\mathrm{03^{h}19^{m}38.99^{s}}$\\
    Dec & $\mathrm{-19^{\circ}23\arcmin42.68\arcsec}$\\
    Explosion epoch & JD 2459918.67$\pm$0.4\\
    Distance & $19.0\pm2.9$ Mpc\\
    $E(B-V)_{MW}$ & $0.024\pm0.001$ mag\\
    $E(B-V)_{host}$ & $0.03\pm 0.01$ mag\\
    $E(B-V)_{total}$ & $0.05\pm 0.01$ mag\\
    Host & NGC 1300 \\
    Redshift & 0.00526\\
\enddata
\end{deluxetable}

We adopt a distance of $19.0\pm2.9$ Mpc from the PHANGS survey \citep{anand_distances_2021} derived using the numerical action method \citep{shaya_action_2017, kourkchi_cosmicflows-3_2020}.

From our Swift observations, we confirmed that there is qualitatively little host extinction.
To quantify the amount of extinction, we obtained medium resolution spectra with Binospec \citep{fabricant_binospec_2019} on MMT and RSS \citep{rangwala_imaging_2008} on SALT. In these spectra the Na I D1 and D2 lines are cleanly separated for both the Milky Way and host galaxy. 
We simultaneously fit the continuum and Gaussian absorption lines to the four Na I D extinction features using the astropy modeling package \citepalias{astropy_collaboration_astropy_2013, astropy_collaboration_astropy_2018, astropy_collaboration_astropy_2022}. 
We calculate the equivalent width using the continuum from this model and the spectrum itself.
Using the relationship of \citet{Poznanski2012} and scaling by 0.86 to match the extinction measurements of \citet{schlafly_measuring_2011}, we measure an average $E(B-V) = 0.021 \pm 0.01$ mag for the \ion{Na}{1} D1 and D2 lines of the Milky Way, confirming the \citet{schlafly_measuring_2011} value of $E(B-V) = 0.026 \pm 0.001$ mag that we obtain from the IPAC Dust Service\footnote{https://irsa.ipac.caltech.edu/applications/DUST/index.html} using the supernova coordinates.
With this agreement, we adopt $E(B-V)_{MW} = 0.026 \pm 0.001$ mag for the remainder of this paper.
For the host galaxy we measure $E(B-V)_{host} = 0.03 \pm 0.01$ mag from the stronger \ion{Na}{1} D2 line, which we will use throughout this paper.

Given that there is less than one day between the last non-detection and first detection, over which time SN~2022acko rose by $>3.3$ mag to near peak magnitude, we define the explosion epoch as the last non-detection. 
We take half the distance between the last non-detection and first detection as the error on the explosion epoch: 0.4 d.

\subsection{Light Curve} \label{sec:lc}
\begin{figure*}%
\centering
\includegraphics[width=1\textwidth]{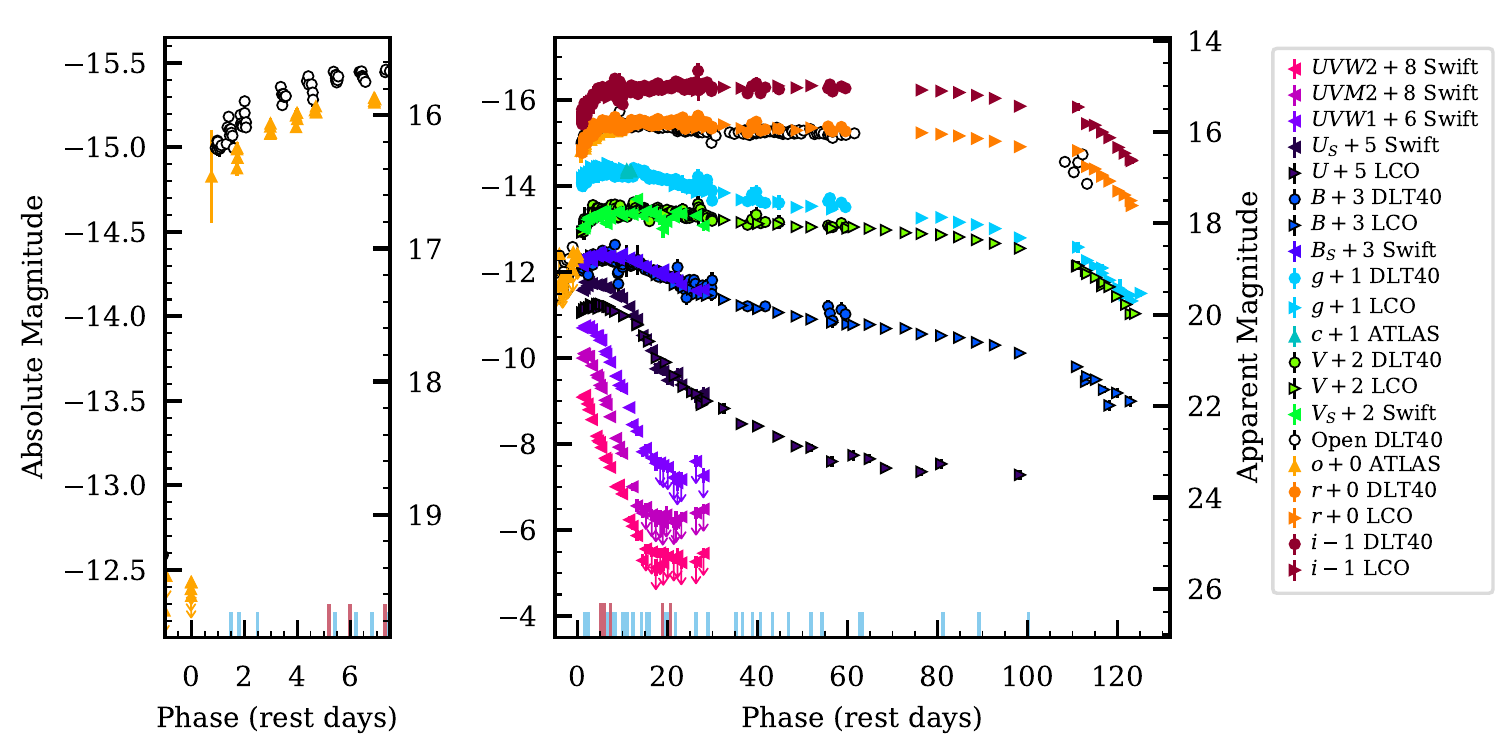}
\caption{The absolute and extinction-corrected apparent magnitude optical and UV light curves of SN~2022acko.
The left panel shows the first 7 days of evolution, highlighting the tightly constraining non-detections within 24 hours of discovery and high-cadence follow up observations obtained immediately by the DLT40 and ATLAS surveys. No offset is applied to these filters in either panel. The DLT40 open filter is calibrated to $r$ band while the ATLAS $o$ filter is similar to $r$+$i$ band (5600-8200~\AA). 
The right panel shows the full multi-band light curve, including the UV evolution. 
Ticks at the bottom of each panel mark epochs of spectroscopic observations with optical spectra marked in light blue and HST UV spectra shown in pink. 
The photometry shown this figure is available as the Data Behind the Figure.}
\label{fig:lightcurve}
\end{figure*}

Given the unique nature of the FUV and NUV spectra of SN~2022acko, we briefly examine its photometric properties and compare it to other well-studied supernovae to provide some context for the UV analysis.
SN~2022acko rose to $V$-band peak brightness $\sim$5 days after explosion and reaching a peak brightness of $V=-15.4$ mag.
\autoref{fig:lightcurve} shows the light curve of SN~2022acko with the left panel highlighting the last non-detection and early rise while the full light curve is shown on the right.

\begin{figure}
    \centering
    \includegraphics[width=1\columnwidth]{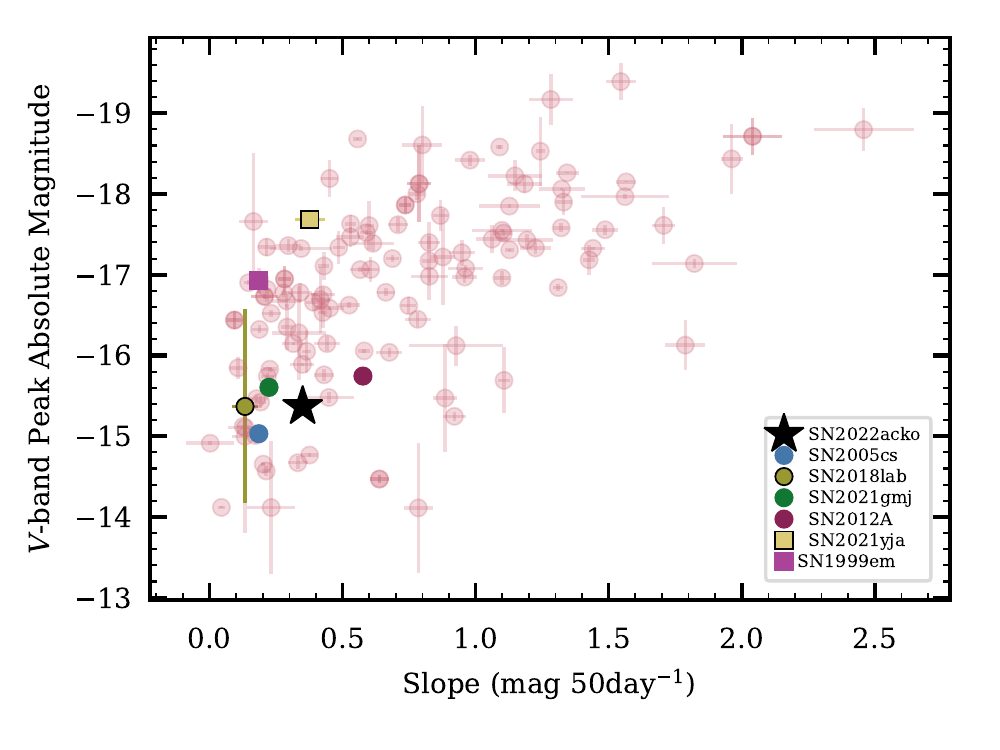}
    \caption{The $V$-band slope at 50 days vs. $V$-band peak absolute magnitude for a sample of Type II supernovae \citep[SNDAVIS database and][]{Valenti2016}, showing that SN~2022acko (black star) is a low luminosity supernova, similar to SN~2005cs (blue circle), SN~2018lab \citep[mustard circle;][]{pearson_circumstellar_2023}, SN~2021gmj (green circle; Meza in prep.), and SN~2012A (magenta circle).
    The slope at 50 days lies between that of SN~2005cs, SN~2018lab,  SN~2021gmj and SN~2012A.
    We include other supernovae with important UV datasets: SN~1999em, and SN~2021yja.}
    \label{fig:PeakvsSlope}
\end{figure}

The light curve evolution is similar to low-luminosity supernovae (LLSNe).
Using the definitions of \citet{Valenti2016} and \citet{anderson_characterizing_2014}, we measure the plateau slope at 50 days  ($s_{50,V}=0.35$ $\mathrm{mag\,(50\,d^{-1}})$) and plateau length ($t_{pt}\sim115$ d).
\autoref{fig:PeakvsSlope} shows the peak absolute magnitude and plateau slope of SN~2022acko compared to a sample of Type II supernovae from \citet{Valenti2016} and our custom database, SNDAVIS.\footnote{https://dark.physics.ucdavis.edu/sndavis/} 
Highlighted are other supernovae of interest: LLSNe: SN~2005cs \citep{brown_early_2007, pastorello_sn_2006, pastorello_sn_2009}, SN~2012A \citep{tomasella_comparison_2013}, SN~2018lab \citep{pearson_circumstellar_2023}, SN~2021gmj (Meza-Retamal et al., in prep.); a normal Type II supernova with early-NUV spectroscopy: SN~2021yja \citep{vasylyev_early-time_2022}; and the only normal Type IIP supernova with FUV spectra: SN~1999em \citep{baron_preliminary_2000}.
Although the peak magnitude is similar to the LLSNe, the plateau slope is slightly steeper and plateau length slightly shorter.
The evolution of SN~2022acko most closely resembles that of SN~2012A in the UV and blue optical and SN~2018lab in the red optical.

The first \textit{uvw2} observation occurs at the peak of the light curve in that filter and the \textit{uvm2} and \textit{uvw1} light curves reach peak brightness at the second datapoint.
It is unusual to catch the UV rise, highlighting how early these observations were obtained \citep{pritchard_bolometric_2014}. 
The UV steadily declines as both the ejecta cool and metal-line blanketing in the UV develops until SN~2022acko is no longer detected above the brightness of the underlying galaxy light (JD 2459933.225: $uvw2$, $uvm2$, JD 2459937.14: $uvw1$).
We continued to monitor it for about a week beyond this, and photometry may be recovered from these observations in the future with template subtraction.

\subsection{Spectroscopy}
\begin{figure}
    \includegraphics[width=1\columnwidth]{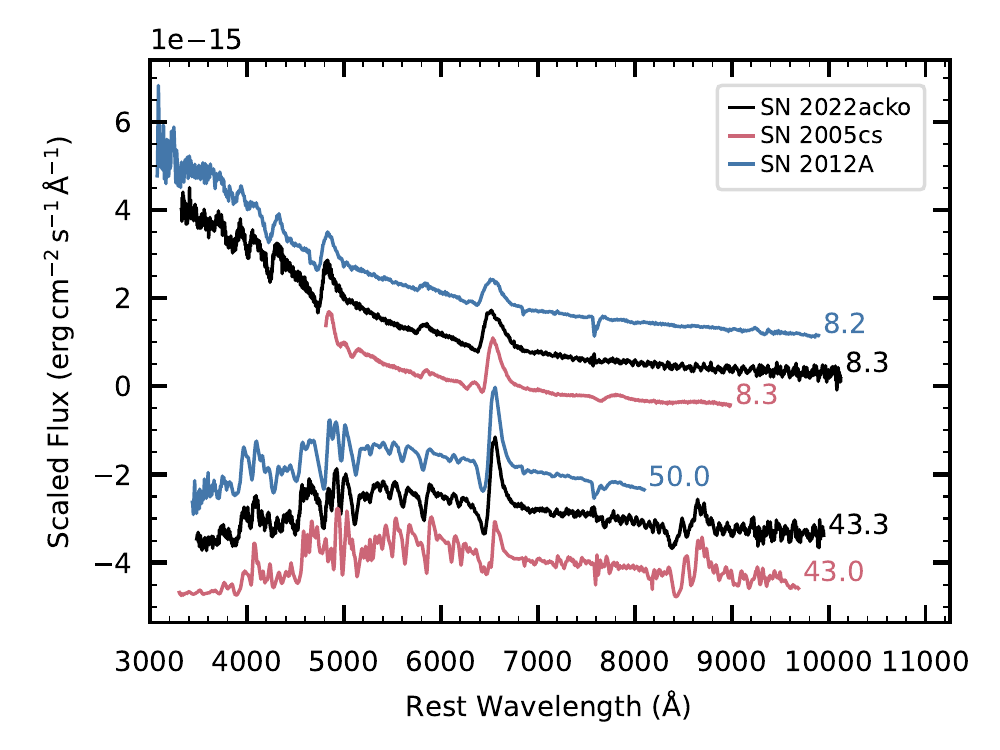}
    \caption{The optical evolution of SN~2022acko (black) is very similar to the LLSNe SN~2005cs \citep[pink;][]{pastorello_sn_2006, pastorello_sn_2009} and transitional luminosity SN~2012A \citep[blue;][]{tomasella_comparison_2013}. 
    The similarity of the spectra can be seen here at both $~\sim$8 and $\sim$50 days post explosion, with SN~2022acko falling between SN~2005cs and SN~2012A in terms of line widths (ejecta velocity).}
    \label{fig:OptSpecComp}
\end{figure}

The first optical spectrum was obtained within 24 hours of discovery and spectroscopic follow-up observations continued at high cadence through the end of the HST observations, with approximately weekly spectra obtained afterwards.
\autoref{fig:OptSpecComp} shows the close resemblance of optical spectra of SN~2022acko to those of SN~2005cs \citep{pastorello_sn_2006} and SN~2012A \citep{tomasella_comparison_2013}, two photometrically similar supernovae.
Consistent with the photometric properties, SN~2012A has slightly broader features, while SN~2005cs has slightly narrower features.
With the exception of \autoref{fig:OptSpecMontage}, the spectra in all figures have been corrected for reddening using the extinction law of \citet{cardelli_relationship_1989} and the extinction values given in \autoref{tab:UVSNComp}.

We note two interesting features in the optical spectra of SN~2022acko. 
First, like many Type IIP supernovae observed at early times, SN~2022acko shows a ledge-like feature in the earliest spectra at 4,600~\AA{} \citep[see ][ and references therein]{pearson_circumstellar_2023}.
Also of note is the brief presence of an absorption feature bluewards of \ion{He}{1} around day 5 which was attributed to both high-velocity \ion{He}{1} \citep{pastorello_sn_2009} and \ion{N}{2}  \citep{baron_preliminary_2000, dessart_quantitative_2005}. 

\section{UV Spectra}\label{sec:uv}
\begin{figure*}
    \includegraphics[width=1\textwidth]{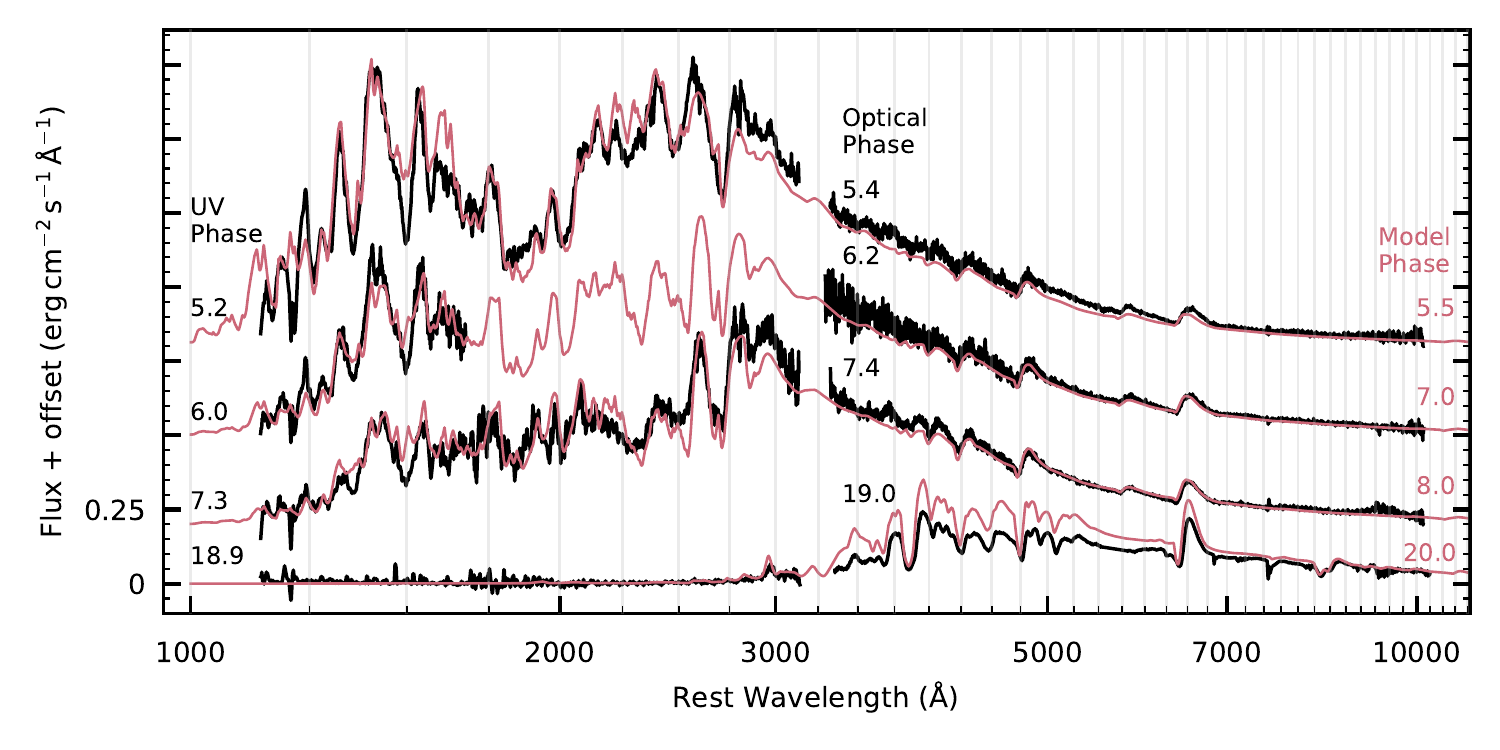}
    \caption{A time series of the extinction corrected UV and optical spectra of SN~2022acko (black) and the well-matched \texttt{CMFGEN} models (pink;  scaled as described in \autoref{sec:CMFGEN}). 
    The phase (in days) of each spectrum is given on the left for the UV observations, in the center for the optical observations, and on the right for the models.
    The day 19 spectra (observed and scaled model) are at their true flux, while the day 7 spectrra are shifted by  $2\times10^{-15}$ \funit, the day 6 spectra is shifted by $5\times10^{-15}$ \funit, and the day 5 spectra is shifted by $8\times10^{-15}$ \funit.
    The model spectra shown in this figure are available as the Data behind the Figure.}
    \label{fig:UVmodelTimeSeries}
\end{figure*}

The UV spectra display a wealth of features which provide key information on the temperature, density, and chemical abundances of the outer ejecta.
This can be seen in the first three UV spectra of SN~2022acko, shown in \autoref{fig:UVmodelTimeSeries}.
However, this information quickly fades as a forest of metal lines blanket the spectra, bringing all FUV flux to nearly zero within the first 20 days.
This UV flux deficit is aggravated by the cooling of the photospheric layers, which causes a strong shift of the SED to longer wavelength until the conditions stabilize at the onset of the hydrogen recombination phase.

Despite the information contained in the FUV and NUV, very little early data exists on Type II supernova.
Two unusual Type II supernovae were studied early with IUE: SN~1980K and SN~1987A \citep{cassatella_spectral_1987, wu_international_1992}. 
SN~1980K showed signs of interaction \citep{pun_ultraviolet_1995} while SN~1987A resulted from a blue supergiant (BSG) progenitor rather than a red supergiant (RSG), leading to a different evolution than expected for normal Type IIP supernovae \citep{walborn_composite_1987, pun_ultraviolet_1995}. 

The UV spectra of SN~2022acko show strong doppler-broadened features in the FUV and NUV which are most prominent in the first spectrum. 
This is the first time most of these features have been observed in the FUV of a Type IIP/L supernova.
The overall flux of the UV spectra fade considerably over the first three HST epochs (three days), most notably between 1100-1800~\AA{} and 2050-2750~\AA{}. 
The final two epochs of UV spectra contain almost zero flux, due to the severe line blanketing, except for a strong feature at 2970~\AA.
This highlights the need for UV observations to be obtained within three weeks of explosion, with earlier observations containing significantly more information.
The spectral evolution can be seen in \autoref{fig:UVmodelTimeSeries}.

Both the interstellar medium (ISM) and CSM can manifest as narrow features in the supernova spectrum \citep{panagia_coordinated_1980, schlegel_sn1987c_1998,leonard_evidence_2000, yaron_confined_2017, fransson_physical_1984}.
UV emission lines can be interpreted as a signature of CSM interaction in Type II supernovae especially when they have a Lorentzian profile with a narrow core and broad wings, indicating optically thick CSM.
However, narrow absorption features can be associated with either optically thin CSM or ISM and are thus more challenging to associate with a supernova unless they are seen to evolve with time.
The UV spectra of SN~2022acko contain a number of narrow absorption features which we list in \autoref{tab:ISM}.
Features which are present at both the host and Milky Way redshifts we associate with ISM.
With these spectra we confirm the host redshift (see \autoref{tab:prop}).
We also note that the galactic and host features are similar in depth, implying that the supernova environment is at approximately solar metallicity, perhaps slightly higher.
Some features are only detected at the host redshift, although we note that these detections are at a lower S/N than the features detected in both galaxies. 
These features have a similarly narrow width to the features identified in both galaxies and for this reason we cannot differentiate between unshocked CSM (from the evolution of the progenitor) and an unusual ISM in the host galaxy.
The lack of symmetric emission lines combined with the presence of broad absorption and P Cygni profiles indicates that at day 5 there is no optically thick CSM.

\begin{deluxetable*}{lccr}
\tablecaption{Narrow absorption features identified in the UV spectra. \label{tab:ISM}}
\tablehead{\colhead{Ion}   & \colhead{Wavelength (\AA)} &  \colhead{Source} & \colhead{Reference}}
\startdata
\ion{Si}{3} & 1206 & galactic, host & \citealt{werk_cos-halos_2013}\\
\ion{Fe}{2}/\ion{Si}{2} & 1260 & host & \citealt{fransson_physical_1984, morton_atomic_2003}\\
\ion{C}{1}/\ion{Si}{2} & 1280 & host & \citealt{fransson_physical_1984}\\
\ion{O}{1}       & 1302, 1304 & host & \citealt{fransson_physical_1984}\\
\ion{C}{2}       & 1335, 1336 & galactic, host & \citealt{fransson_physical_1984}\\
\ion{N}{2}             & 1344 & galactic, host & \citealt{fransson_physical_1984}\\
\ion{Si}{4}      & 1394, 1403 & galactic, host & \citealt{fransson_physical_1984}\\
\ion{Si}{2}            & 1527 & galactic, host & \citealt{fransson_physical_1984}\\
\ion{C}{4}       & 1548, 1551 & galactic, host & \citealt{fransson_physical_1984}\\
\ion{C}{1}       & 1560, 1561 & host & \citealt{fransson_physical_1984}\\
\ion{Fe}{2}            & 1589 & host & \citealt{morton_atomic_2003}\\ 
\ion{Fe}{2}            & 1608 & galactic, host & \citealt{morton_atomic_2003}\\ 
\ion{He}{2}            & 1640 & host & \citealt{fransson_physical_1984}\\
\ion{Fe}{2}           & 2344 & galactic, host & \citealt{morton_atomic_2003}\\
\ion{Fe}{2}           & 2383 & galactic, host & \citealt{morton_atomic_2003}\\
\ion{Fe}{2}           & 2600 & galactic, host & \citealt{cardelli_atomic_1995}\\
\ion{Mg}{2}     & 2796, 2802 & galactic, host & \citealt{panagia_coordinated_1980}\\
\enddata
\end{deluxetable*}

\subsection{Radiative-Transfer Modeling}\label{sec:CMFGEN}
Unlike the optical, early UV spectra are heavily affected by metal lines, producing a complex spectrum that strongly deviates from a blackbody. 
To identify the elements present and characterize the ejecta properties such as density and temperature we model the full UV+optical spectral evolution with the non-local thermodynamic equilibrium (NLTE) radiative-transfer code CMFGEN \citep{hillier_treatment_1998, hillier_time-dependent_2012, hillier_photometric_2019, dessart_type_2013}.

We start with a zero-age main sequence 12 \msun{} star at solar metallicity ($Z=0.014$), evolved in Modules for Experiments in Stellar Astrophysics \citep[\texttt{MESA};][]{,paxton_modules_2013, paxton_modules_2015, paxton_modules_2018}  with default parameters and a dutch wind scaling factor of 0.8. 
We chose a relatively low-mass progenitor as these are associated with under-luminous events \citep[e.g. SN~2008bk][]{maund_new_2014}, however, we do not place too much emphasis on the progenitor mass as the effects of progenitor mass are degenerate with a number of other uncertain physical processes in stellar evolution (see discussion later in the section).
The pre-explosion progenitor has a final mass of 9.7 \msun{}, an effective temperature of 4002~K, a luminosity of 57802~\lsun{}, and a radius of 500~\rsun{}.
Its envelope composition is $M(\mathrm{H})=4.44$~\msun{}, $M(\mathrm{He})=3.15$~\msun{}, and $M(\mathrm{O})=0.27$~\msun{}. 
\texttt{MESA} artificially cuts the density profile at a Rosseland mean optical depth of 2/3 producing a dense surface ($\rho\sim1\times10^{-9}\, \mathrm{g \,cm^{-3}}$), which is unsuitable for modeling the shock breakout phase. 
While material beyond this point is not important during stellar evolution, the supernova shock ionizes it, causing it to become optically thick. We extrapolate a steep density gradient beyond the surface of the original \texttt{MESA} model down to a density of $\rho < 1\times10^{-12}\, \mathrm{g\, cm^{-3}}$, including a total of 0.02~\msun{} of additional material.

A thermal bomb is used to explode the model using the radiation hydrodynamics code \texttt{V1D} \citep{livne_implicit_1993, dessart_determining_2010, dessart_shock-heating_2010} depositing $7.4\times10^{50}~\mathrm{erg}$ for 0.1s between mass shells 1.55 and 1.60 \msun{} in the progenitor, producing an ejecta kinetic energy $6\times10^{50}~\mathrm{erg}$ and an ejecta mass of 8.16~\msun{}.
The mass cut at 1.55 \msun{} corresponds to the neutron star gravitational mass. 
At day 5, the ejecta are mapped to CMFGEN, which performs the time-dependent radiation transport in NLTE assuming homologous expansion.
Model ages are derived from the homologous expansion in the outer ejecta layers, neglecting the stellar radius, but assuming constant velocity since t=0, with $t=R/V$ (where R/V is set using a mass shell at 7800 $\rm km\,s^{-1}$).
This supernova age is thus in tension with the explosion data, compounding with the uncertainty associated with the $\sim 1$ day it takes the shock to reach the surface of the RSG progenitor.
Comparing our homology evolution with the evolution of the model using the radiation hydrodynamics code \texttt{V1D}, we find agreement to within 5\% for both the photospheric radius and velocity through day 20.
This confirms that our assumption of homology at 5d in CMFGEN is good and could be used to model a variety of different progenitor parameters such as metallicity, initial mass, and explosion energy.
As this model was created specifically to characterize the early data presented in this paper, we ignore a number of effects which are only important at later times when the photosphere begins to recede deep into the hydrogen-rich ejecta layers.
First, we truncate the model at a minimum velocity of 2000~$\mathrm{km\,s^{-1}}$, excluding the deep and very optically thick inner ejecta.
We also do not include $\mathrm{{}^{56}Ni}$ and do not treat the non-thermal effects that arise from radioactive decay.

Ideally, a model with optimized parameters would reproduce the observed flux from the UV through the IR.
For this analysis we did not attempt to optimize parameters, instead using an existing model with an explosion energy of $E_{\mathrm{kin}}=6\times10^{50}~\mathrm{erg}$.
With this model, we find that, without the scaling described and shown below, the model overestimates the UV flux by a factor of two. 
If this were attributed to an error in distance, NGC~1300 would have to be at 26.8 Mpc, close to $3\sigma$ above our distance estimate.
The overall flux level of a supernova is affected by the explosion energy, progenitor radius, and contribution from CSM. 
While a full exploration of this parameter space is beyond the scope of this paper, we explored a model with lower explosion energy ($\sim4\times10^{50}\mathrm{erg}$). 
However, we found that while the overall flux of the lower energy model was a better match to the observed flux, the line widths were too narrow and the optical flux, even at earlier times, was brighter than the observed flux when the model spectra were scaled to match the UV flux with the same method as the higher energy model.
We therefore suggest that future modeling explore different progenitor radii and CSM configurations.
We note that due to degeneracies in the modeling parameters, a smaller radius could be achieved with either a lower mass progenitor or a progenitor of the same mass but with more vigorous convection in the hydrogen envelope.

Given the mismatch between the model flux and observed flux, we scale our model flux in the following way.
We redshift and redden the model using the redshift and extinction of SN~2022acko.
We then scale the first model to the Swift UV photometry ($uvw2$, $uvm2$, and $uvw1$) at 5.5 days using a constant offset. 
Next, we return the model to rest wavelengths and unextinguished flux.
We repeat this process on all other model spectra, applying the scaling factor derived from the first spectrum to preserve the flux evolution of the models. 
The observed spectra are then compared to the model at a similar phase.
\autoref{fig:UVmodelTimeSeries} shows the results of our modeling.
It is striking how well the features and flux evolution are reproduced, implying that this model is a good representation of the outer ejecta velocity, temperature, and density.
Additionally, the similarity of the model and observed spectra imply that the supernova is approximately solar metallicity \citep[see][for a description of the effects of metallicity on UV spectra]{dessart_type_2013}.
Given that the scaling is set by the first UV spectrum, the model reproduces the flux evolution very well in the UV, including the final epoch at day 18.9.
In the optical, it slightly underestimates the flux in the early observations.
It is possible that this could be solved at early times with the introduction of clumping into the model which would more rapidly shift the flux from the UV to the optical \citep{dessart_impact_2018}, or by changing the progenitor radius, which can alter both the absolute and relative flux distribution \citep{dessart_type_2013}.
With more optimization, a fully consistent model could be found.

Alternately, it is possible that the host extinction is underestimated.
We find an additional extinction of $E(B-V)=0.03$ mag produces a better fit to the first three epochs.
However, the final optical spectrum at day 18.9 is overestimated by the model, which is made worse when the additional extinction is added.
This could be mitigated by invoking some dust very close to the progenitor which is destroyed between days 7 and 19 by the interaction with the supernova ejecta \citep{fraser_red_2012, kochanek_absorption_2012, van_dyk_red_2012}. 
However, even with no extinction correction, the model is too bright at later phases, indicating that time-dependent extinction cannot fully account for the discrepancy.
Although HST pre-explosion observations exist \citep{van_dyk_identifying_2023}, the photometry does not constrain this mild level of extinction.
Nevertheless, the similarity of the model spectra to the observed spectra at both optical and UV wavelengths on days 5-7, indicates that there is either weak or no interaction at this phase and that the dominant radiation source is the fast moving ejecta.

As a consistency check, we perform synthetic photometry on the scaled CMFGEN model spectra to generate a synthetic light curve which we compared to the observed light curve through day 25.
The trends as a function of wavelength seen in the spectroscopic comparison in \autoref{fig:UVmodelTimeSeries} are also apparent in the light curve evolution.
Although the scaling is set to match the first model epoch to the observed light curve in the Swift $uvw2$, $uvm2$, and $uvw1$ filters on day 5, the full UV evolution is very well matched by the models, including the point at which the supernova drops below detection.
The optical light curve rises more slowly than the observed light curve, reaching $V$-band peak about 5 days after the observed light curve.
Given that the model light curve does not start until day 5, when the observed light curve has flattened from its initial rise, the discrepancy between the observed and model flux is never too great.
The $g$, $V$, $r$, and $i$-band peaks are brighter than the observed peaks by 0.1-0.3 mag.
As with the spectra, these discrepancies may be resolved by running a more customized model that varies either the progenitor radius or CSM around the progenitor.

\begin{figure}
    \includegraphics[width=1\columnwidth]{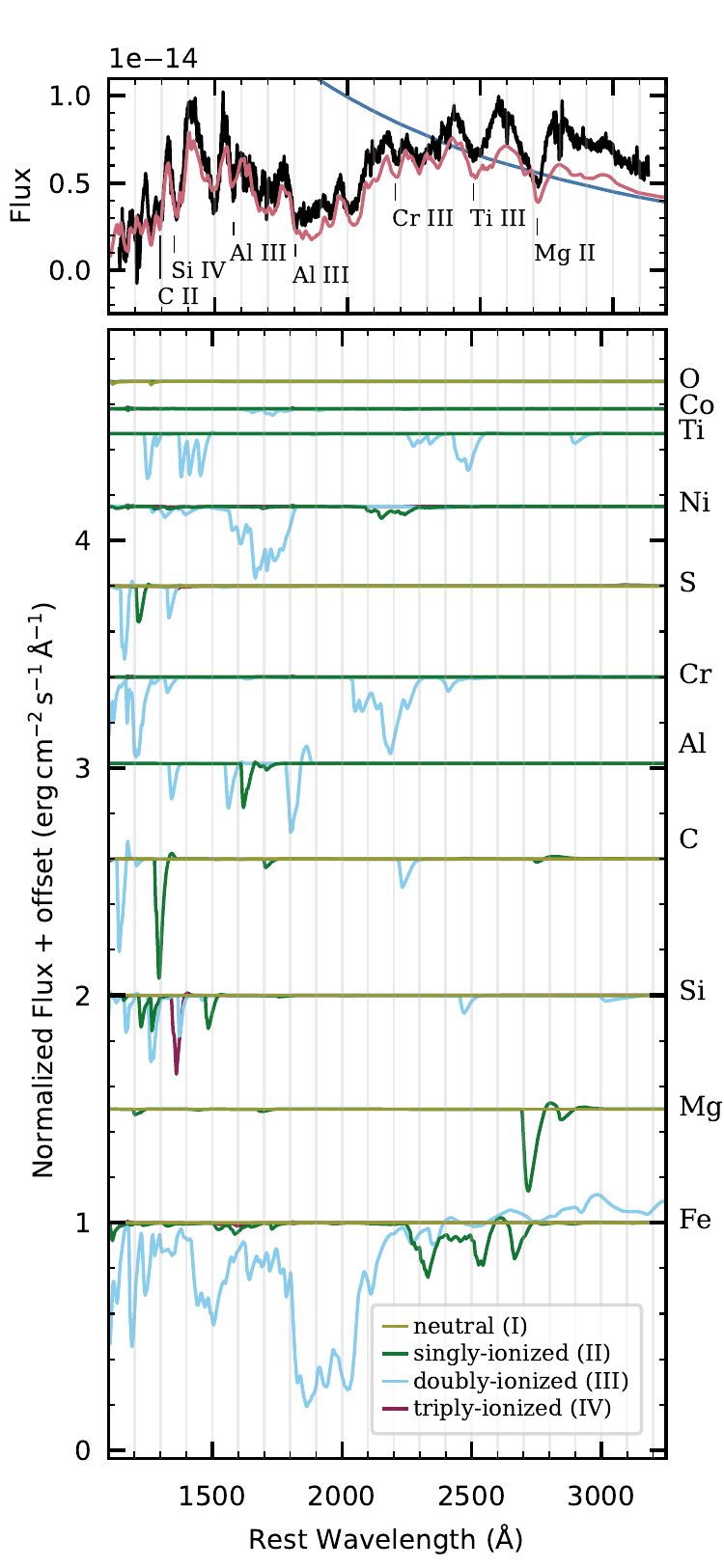}
    \caption{The identification of species present in the day five HST/STIS UV spectrum of SN~2022acko. 
    \textit{Top}: The UV spectrum of SN~2022acko (black) with the total \texttt{CMFGEN} low energy model overplotted in pink.
    The \texttt{CMFGEN} continuum flux is shown in blue, demonstrating the strong deviation of the observed spectrum from a blackbody.
    Although this is a complicated region, a few fairly isolated lines are identified below the spectrum.
    \textit{Bottom}: The contribution of the most prominent ions, grouped by species, to the full spectrum. 
    The model spectra of individual ions in this figure are available as the Data behind the Figure.}
    \label{fig:UVmodelLadder}
\end{figure}

In \autoref{fig:UVmodelLadder} we show the spectroscopic contributions of the most prominent elements in the UV. 
Individual ion spectra are calculated using CMFGEN, omitting their bound-bound transitions from the formal solution to the radiative transfer equation.
The ion spectrum is then found by taking the ratio of the full spectrum to that with the ion omitted.
The spectrum is most strongly influenced by \ion{Fe}{3} at these epochs, which produces a complex structure of absorption lines, fundamentally altering the continuum. 
Beyond \ion{Fe}{3}, we identify significant contributions from \ion{Fe}{2}, \ion{C}{3}, \ion{C}{2}, \ion{Ti}{3}, \ion{Si}{4}, \ion{Si}{3}, \ion{Si}{2}, \ion{S}{3}, \ion{S}{2}, \ion{Ni}{3}, \ion{Cr}{3}, \ion{Al}{3}, \ion{Al}{2}, and \ion{Mg}{2}.
The strongest contributions from C and Si are in the FUV only and are clearly identified in our spectra.
We label the relatively isolated lines in the top panel of \autoref{fig:UVmodelLadder}.
As has been noted by other authors, the UV spectrum is not a continuum with emission or absorption lines but rather a continuous set of features which are blended together making it challenging to identify individual components \citep[e.g.][]{pun_ultraviolet_1995, gezari_probing_2008, dessart_using_2008, dessart_supernova_2010}.
With these models, we also find that the 2970~\AA{} feature in the day 19 and 20 spectra is not emission but rather a window of lower absorption from the nearby \ion{Fe}{2}, \ion{Cr}{2}, and \ion{Ti}{2} absorption complexes.

\begin{figure*}
    \includegraphics[width=1\textwidth]{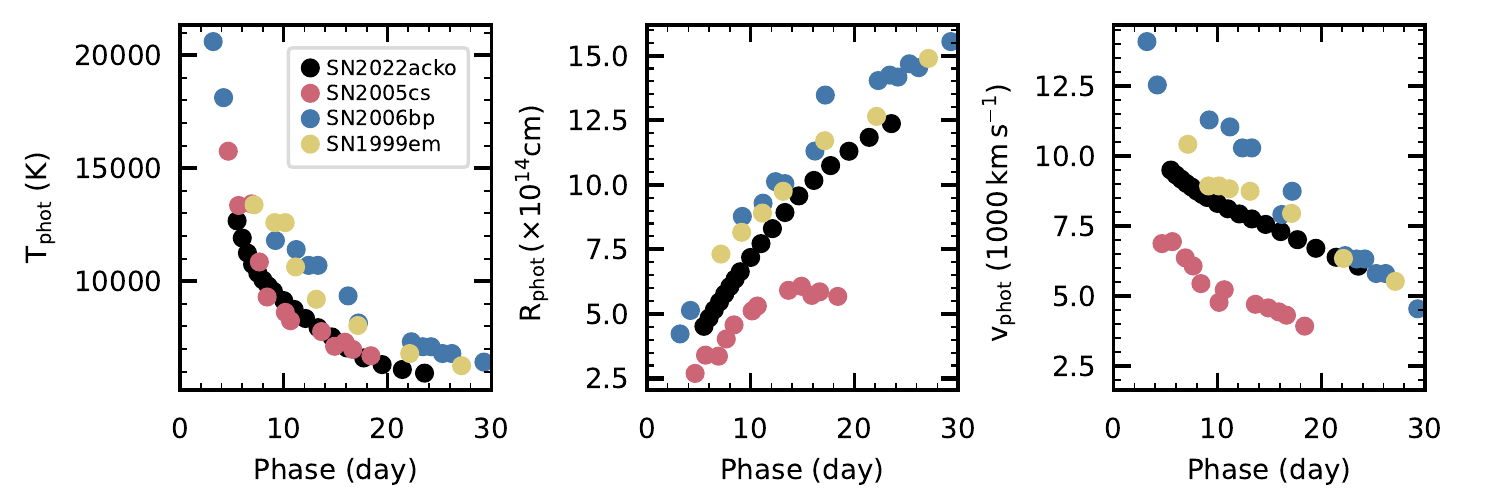}
    \caption{The photospheric evolution of SN~2022acko compared to those of SN~1999em \citep{dessart_quantitative_2006}, SN~2005cs, and SN~2006bp \citep{dessart_using_2008}.
    The photosperic temperature (left), radius (middle), and velocity (right) fall between the derived parameters for SN~1999em, SN~2005cs, and SN~2006bp, indicating that this is not an unusual event.}
    \label{fig:profile}
\end{figure*}

The ejecta radius, temperature, and velocity influence the spectroscopic properties of a supernova.
Although conditions vary throughout the ejecta, we measure these quantities at the photosphere, allowing us to compare the ejecta of different supernovae.
We extract these parameters for SN~2022acko from the \texttt{CMFGEN} model at the photosphere and compare them with other publicly available models (SN~2005cs and SN~2006bp: \citealt{dessart_using_2008} and SN~1999em: \citealt{dessart_quantitative_2006}).
Unlike the model for SN~2022acko, the models of SN~1999em, SN~2005cs, and SN~2006bp were computed using a steady-state version of \texttt{CMFGEN} which fit the observed spectra by varying the supernova luminosity, radius, and velocity. 
While velocity and temperature are well constrained by the ions present in the spectra and the widths of the spectral features, there is a degeneracy between the luminosity and radius which makes these values unreliable in these older models.
For this reason, we do not use the original photospheric radius of the model. 
Instead, we derive an accurate photospheric radius using $R_{\mathrm{phot}}=D_{EPM}\times\theta_{EPM}$, where $\theta_{EMP}$ is the angular size of the photosphere and $D_{EPM}$ is the distance to the supernova, both derived as part of the expanding photosphere method, and a set of multi-epoch steady-state \texttt{CMFGEN} calculations \citep[with method described in][] {dessart_quantitative_2005}.
The results are shown in \autoref{fig:profile}.
The photospheric temperature, radius, and velocity all fall within the range of other Type II supernovae.
The photospheric temperature is very similar to SN~2005cs: starting at $\sim12700$ K on day 5.5 and decreasing to $\sim6000$ K by day 23.5, indicating the onset of hydrogen recombination.
This implies that SN~1999em and SN~2006bp are emitting more UV flux relative to the optical flux, although the different temperature will also affect the strength of the line-blanketing.
The photospheric velocity is slightly lower than SN~1999em and SN~2006bp, but significantly higher than SN~2005cs at early times, converging to a similar velocity around day 20.
The velocity decreases from $\sim9500$ $\mathrm{km\,s^{-1}}$ on day 5.5 to $\sim6000$ $\mathrm{km\,s^{-1}}$ on day 23.
Similarly, the photospheric radius for SN~2022acko falls between that of SN~2006bp and SN~2005cs at early times, starting with a radius of $4.5\times10^{14}$ cm ($\sim6500$ \rsun{}).
It remains between these supernovae for its full evolution, although it does approach the photospheric radius of  SN~1999em by day 23.5 with $R_{\mathrm{phot}}=12.4\times10^{14}$ cm ($\sim18000$ \rsun{}).
These values are tabulated in \autoref{tab:PhotProp}.

The light curve and spectroscopic properties of a supernova depend sensitively on the physics of stellar evolution and supernova explosions.
Processes like convection (whether at the surface or the core) are not directly observable, are too complicated to model from first principles, and computationally intensive to simulate with full physics in 3D \citep{goldberg_numerical_2022}. 
We are therefore still at the stage of constraining fundamental parameters, such as progenitor radius, by creating models that capture a supernova's evolution from explosion through nebular phase at all wavelengths.
One of the most comprehensive explorations of the effects of different stellar evolution and supernova explosion parameters was performed by \citet{dessart_type_2013}.
Using \texttt{MESA}, \texttt{V1D}, and \texttt{CMFGEN} they systematically varied one parameter at a time to explore the effects of progenitor radius (via the mixing length parameter), core overshooting, rotation, metallicity, mass-loss, and explosion energy.
They find that the progenitor radius can affect the light curve rise time, luminosity, and color evolution.
Progenitor metallicity, which is unaltered in the outer layers of the ejecta by explosive nucleosynthesis, affects nuclear reactions, progenitor radius, and mass-loss, affecting both the light curve shape and brightness, as well as spectroscopic features, most strongly in the UV.
Unfortunately, rotation and core-overshooting mimic increasing progenitor mass, creating larger He core masses, higher luminosities, higher mass-loss rates, and larger radii.
While these effects are measurable, this degeneracy makes it difficult to constrain any of these parameters \citep{dessart_difficulty_2019}.
Increasing explosion energy creates a brighter and shorter plateau, increases spectral line widths, and creates more ${}^{56}Ni$, which creates a brighter radioactive tail phase.
Although we note that there is some degeneracy here as increasing  ${}^{56}Ni$ mass increases the plateau length \citep{Kasen2009, goldberg_inferring_2019}.
In a separate study, \citet{dessart_modeling_2022}, find that depending on density, CSM can have a range of effects: narrow emission lines, high velocity absorption features, increased UV flux, and shallow P Cygni profiles for lower density CSM.
The success of our model in the UV, indicates that the modeling of gas properties and the atomic data are satisfactory and the replication of the majority of spectroscopic features, especially the shape of the UV flux, indicates that the metallicity we selected is a good match for our observations.
Given that these are the first FUV observations at this phase and the wealth of features present, this was in no way guaranteed, and is a validation of the assumptions in the models.
This work also validates the assumption of homology as early as 5 days after explosion, earlier than this assumption has been tested (previous work started 10-15 days after explosion), opening the door to further models at earlier times, when CSM interaction is most likely and the RSG mass loss that produces it least understood.
However, the discrepant luminosity indicates that the explosion energy, radius, or CSM (or some combination) is not quite right in the model.
We attempted to reduce the explosion energy but this resulted in too narrow line profiles compared to observations. 
Given the color dependence of this trend and the fact that CSM would increase the UV luminosity, requiring a lower explosion energy, we conclude that the most likely resolution would be a decrease in progenitor radius, although this is not a possibility that we have yet explored.
A proper grid of early-time models for different progenitor masses, explosion energy, and metallicity is needed.

\subsection{Comparison to other early UV spectra}\label{subsec:comparison}
\begin{figure*}
    \includegraphics[width=1\textwidth]{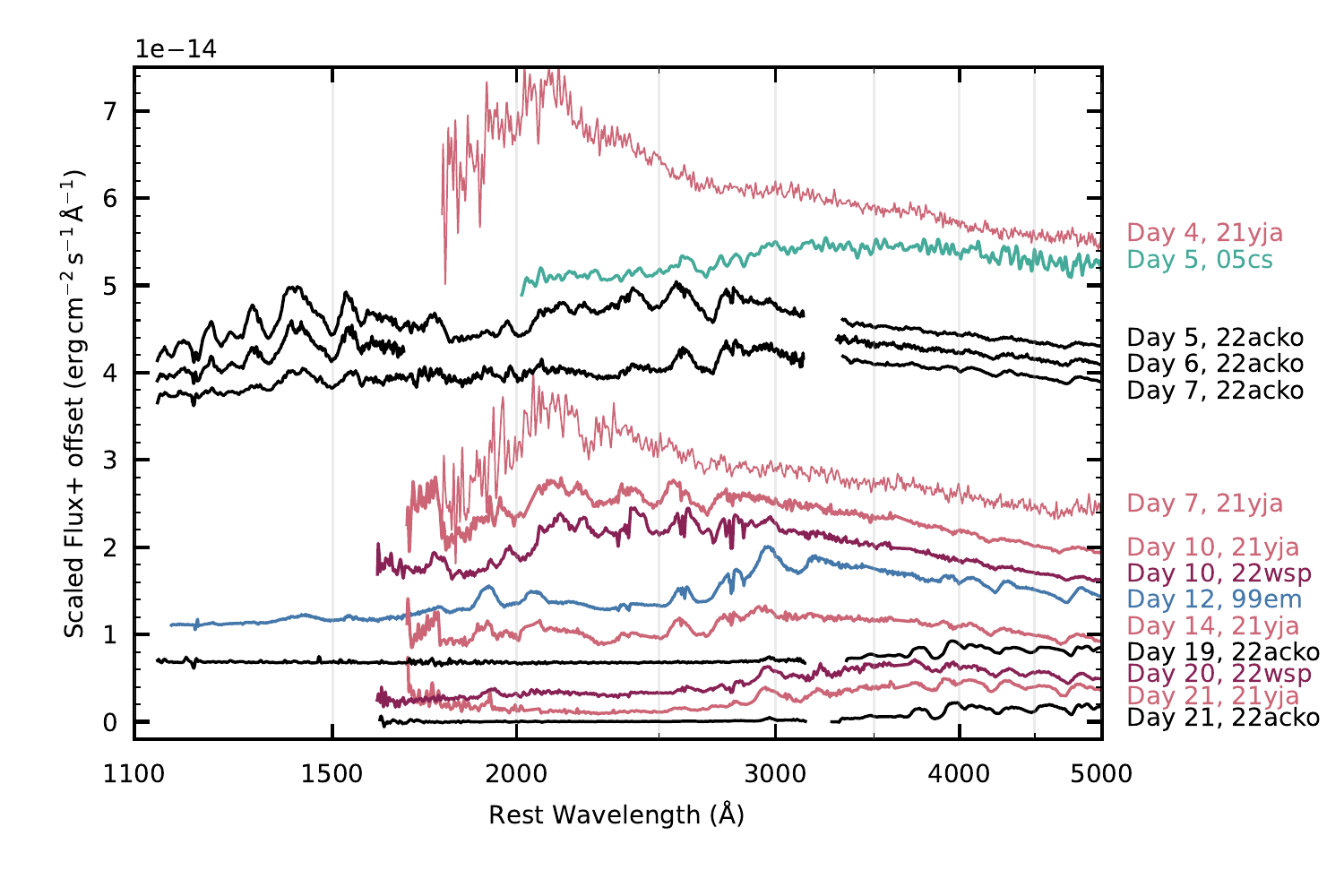}
    \caption{SN~2022acko (black) in the context of the UV evolution of Type IIP supernovae, showing the diversity of UV spectra. 
    Swift and HST NUV spectra of SN~2021yja \citep[pink;][]{vasylyev_early-time_2022} span 4-21 days post explosion.
    HST NUV spectra of SN~2022wsp (maroon) are shown for day 10 and 20 \citep{vasylyev_early-time_2023}.
    A Swift spectrum of the LLSN SN~2005cs (teal; Rowe in prep.) is shown as well.
    SN~2022acko and SN~1999em \citep[blue;][]{baron_preliminary_2000} are the only Type IIP/L supernovae with FUV spectra, with the first epochs of SN~2022acko observed significantly earlier than SN~1999em, resulting in strong FUV features.
    Iron line blanketing is already significantly suppressing the FUV flux by day $\sim$12 in the SN~1999em spectrum.}
    \label{fig:UVSpecComp}
\end{figure*}

The number of Type IIP/L supernovae observed in the UV is very limited making it challenging to compare spectra of different supernovae at the same epoch. 
However, combining observations from different supernovae, it is possible to create a time-series showing the UV evolution (see \autoref{fig:UVSpecComp}).

The only FUV observation of a normal Type II supernova is SN~1999em \citep{baron_preliminary_2000} which was observed $\sim$12 days post explosion and shows a significant, potentially abnormally large, flux deficit in the FUV.
Two other supernovae were well-studied in the NUV by Swift are SN~2021yja \citep{vasylyev_early-time_2022} and SN~2005cs \citep{brown_early_2007}.
The earliest data for both of these supernovae was taken  at four and five days post-explosion, respectively, and have relatively low signal-to-noise (S/N).
Both of these supernovae continued to be monitored by Swift.
A similar dataset was collected in the NUV by GALEX for SN~2005ay, with observations starting 9 days after explosion \citep{gal-yam_galex_2008}.
SN~2021yja was observed with HST ten days after explosion, showing similar lines as the NUV spectra of SN~2022acko \citep{vasylyev_early-time_2022}.
Additional HST NUV spectra of SN~2021yja were obtained on days 14 and 21, showing the suppression of the UV flux with time.
A similar NUV spectroscopic sequence was obtained for SN~2022wsp \citep{vasylyev_early-time_2023} with HST spectra observed on days 10 and 20.
Early HST data of SN~2020fqv were also obtained, although due to high extinction, there is very little flux in the NUV observations and no flux in the FUV observation \citep{tinyanont_progenitor_2022}.

We select the best spectra from this set of Type IIP/L supernovae to create the time series shown in \autoref{fig:UVSpecComp} using our spectra of SN~2022acko, supplemented with SN~2022wsp, SN~2021yja, SN~2005cs, and SN~1999em. 
For each supernova we correct for redshift and extinction and then scale the overall flux level to be approximately that of SN~2022acko at a similar phase using the values in \autoref{tab:UVSNComp}. 
For supernovae that were not observed at a comparable phase, we used the already scaled spectra of SN~2021yja.
While the strong Mg II ($\mathrm{\lambda 2800}$~\AA) feature is visible in many spectra, all observations before day 9 lack high-enough S/N to distinguish other spectral features, with the exception of SN~2022acko.
Additionally, from the strong features visible in SN~2022acko, it is clear that the day 12 FUV spectrum of SN~1999em already suffers from significant iron line-blanketing and the shape of the NUV spectrum is strikingly different from the other supernovae observed at similar epochs.
We note that the explosion epoch of SN~1999em is uncertain with 9 days between the last non-detection and first detection.
The oldest the FUV spectrum of SN~1999em could be is 16 days, shifting it down one spectrum in \autoref{fig:UVSpecComp}. 
We tested shifting and scaling the spectrum of SN~1999em to match SN~2021yja at day 14, SN~2022acko at day 19, and SN~2022wsp at day 20. 
We find that while the UV and optical can individually be well-matched to SN~2021yja at day 14, the UV flux is smaller compared to the optical for SN~1999em.
With further scaling, we find that the UV flux of SN~1999em is too high relative to the later time spectra. 
Both of these results are consistent with an age of 16 days, however, no UV spectrum exists at this phase with which to compare. 
Interestingly, this age is consistent with the age derived by \citet{dessart_quantitative_2006} using the EPM method with the filter and phase combination that produces an EPM distance to the host galaxy of SN~1999em (NGC~1637)  that matches the Cepheid distance \citep{leonard_cepheid_2003}.
The compilation of these UV spectra does not show the uniformity suggested by \citet{gal-yam_galex_2008}. 
In addition to the differences noted above for SN~1999em, SN~2021yja, and SN~2022wsp continue to show NUV flux out to 20 days, possibly due to weak interaction \citep{hosseinzadeh_weak_2022, kozyreva_circumstellar_2022, vasylyev_early-time_2023, dessart_modeling_2022}. 

\subsection{Missing UV Flux Fraction}
\begin{figure}
    \centering
    \includegraphics[width=\columnwidth]{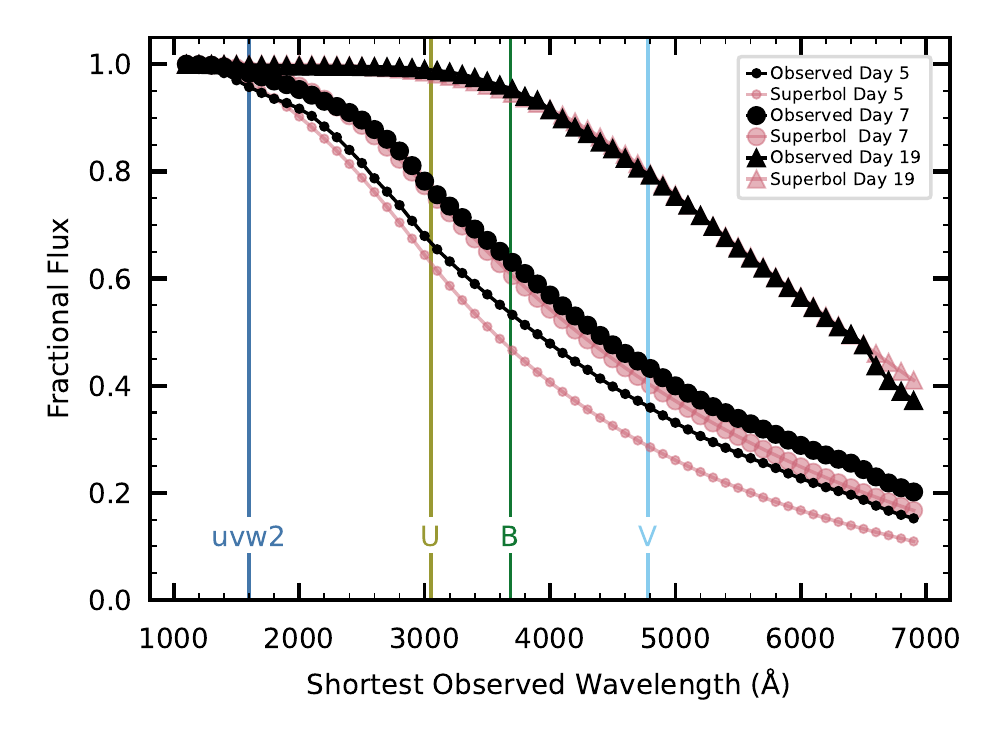}
    \caption{The observed fractional throughput (black) as a function of shortest wavelength observed for day 5 (dot), day 7 (circle), and day 19 (triangle) spectra. 
    Plotted in pink in the fractional flux from the best fit absorbed blackbody spectrum calculated using \autoref{eqn:superbol} and with the same symbols representing the different epochs as were used for the observed flux.
    At all epochs, the absorbed blackbody approximation yields fractional flux within 7\% of the true value.
    The cutoff wavelengths that were used for the filters given in \autoref{tab:frac_tp} are shown as vertical lines and labeled on the plot.}
    \label{fig:frac_tp}
\end{figure}

While Type II supernovae peak in the UV at early times, the majority of our observations are in the optical. 
To calculate a bolometric luminosity, the UV contribution must be inferred from optical observations.
However, as our observations show (see also \citealt{dessart_quantitative_2005, vogl_spectral_2019}), a blackbody spectrum is a poor approximation of the UV flux, even as early as 5 days after explosion (\citealt{dessart_determining_2010} showed that the UV spectrum approaches a blackbody spectrum at photospheric temperatures of 30000~K, well above the temperature of our earliest spectrum). 
Even when Swift observations are available, the $uvw2$ filter cuts off at 1600~\AA.
We calculate the fraction of flux redwards of a given wavelength by combining our observed UV and optical spectra (linearly interpolating across the gap around 3000~\AA), integrating the flux redwards of a given wavelength, and comparing that to the integrated flux over the full wavelength range.
The day 5 and day 7 UV spectra have been scaled (with a small uniform offset) to the UV photometry, while the final day 19 spectrum has not been scaled as the supernova is not detected in the Swift UV at this epoch.
The optical spectra are scaled to the $gri$- photometry with a linear scaling.
As our spectra only span 1150-10150~\AA, this analysis ignores contributions at longer and shorter wavelengths.
\autoref{fig:frac_tp} shows the fraction of flux as a function of shortest wavelength observed from 1100-7000 \AA{} for each epoch.
We also mark the shortest wavelength in the $uvw2$, $U$, $B$, and $V$ filters, indicating the fraction of flux lost when a given filter is the bluest observed.
The fractional flux by filter is tabluated in \autoref{tab:frac_tp}.

Over time, as the peak of the SED shifts to the optical, the majority of the flux is captured with filters accessible from ground-based observatories. 
We find the majority of the flux is captured when the bluest filter is the Swift $uvw2$ filter. 
However, only 53\% of the flux is captured at day 5 if the bluest filter is the $B$-band. 
This fraction of observed flux increases to 63\% two days later as the SED peak shift redwards and is at 95\% by day 19.

\begin{deluxetable*}{lccccr}
\tablecaption{Fraction of flux redward of a given wavelength\label{tab:frac_tp}}
\tablehead{\colhead{Shortest Filter}   & \colhead{Wavelength Range (\AA)} & \colhead{Fraction of Flux (day 5)} & \colhead{Fraction of Flux (day 7)} & \colhead{Fraction of Flux (day 19)}}
\startdata
$uvw2$  & 1600--10150  & 0.96 & 0.98 & 1.00\\ 
$U$     &  3050--10150 & 0.67 & 0.76 & 0.99\\ 
$B$     & 3685--10150  & 0.54 & 0.63 & 0.95\\ 
$V$     & 4780--10150  & 0.36 & 0.44 & 0.79\\ 
\enddata
\end{deluxetable*}

We characterize the observed spectra as a blackbody which undergoes absorption as a function of wavelength, for wavelengths shortward of a cutoff wavelength as described in \citet{nicholl_magnetar_2017} and implemented in the code \texttt{SuperBol} \citep{nicholl_superbol_2018, superbol}.
Briefly, we fit the equation:
\begin{equation}\label{eqn:superbol}
F_{\lambda}(\lambda) = 
    \begin{cases}
    A \times B_{\lambda}(T) & \lambda > \lambda_{\mathrm{cutoff}} \\
    A \times B_{\lambda}(T)\times ( \frac{\lambda}{\lambda_{\mathrm{cutoff}}} ) ^{x} & \lambda \leq \lambda_{\mathrm{cutoff}} \\
    \end{cases}
\end{equation}
Where A is a scale factor, $B_{\lambda}$ is the Plank function, T is the temperature in K, and $\lambda_{\mathrm{cutoff}}$ is the cutoff wavelength in angstroms.
We find the spectra are well approximated by this functional form with the best fit parameters for each epoch given in \autoref{tab:superbol}.
It is interesting that the temperatures derived from this method are significantly higher than those of our \texttt{CMFGEN} model, indicating that while this is a good empirical description of the spectra, it may not be appropriate for characterizing the physical conditions of the ejecta, especially at early times.
A comparison of the observed fractional flux and fractional flux from the analytic \texttt{SuperBol} parameterization is shown in \autoref{fig:frac_tp}.
At early times, the absorbed blackbody approximation underestimates the fractional throughput with a maximum difference of $\sim$7\%, with the effect decreasing over time and as the shortest wavelength observed moves redwards. 

\begin{deluxetable*}{lccccr}
\tablecaption{Absorbed blackbody parameters \label{tab:superbol}}
\tablehead{\colhead{Phase}   & \colhead{Temperature (K)} & \colhead{x} & \colhead{$\lambda_{\mathrm{cutoff}}$} & \colhead{Scale Factor}}
\startdata
5  & 21750  & 2.3& 2960 &  $3.5\times10^{-31}$\\ 
7  &  14150 & 2.0 & 2990 & $9.9\times10^{-31}$\\ 
19 & 7425   & 2.54 & 4815 & $6.2\times10^{-30}$\\ 
\enddata
\end{deluxetable*}

\section{Conclusion}\label{sec:conclusion}
In this paper we present the first early FUV spectra of a Type IIP/L supernova. 
SN~2022acko is similar to SN~2012A, SN~2005cs, SN~2018lab, and SN~2021gmj with a faint $V$-band peak magnitude of $V$=$-$15.4 mag and plateau slope of $s_{50,V} = 0.35~ \mathrm{mag\,(50d^{-1})}$. 
Discovered within 24 hours of explosion, we obtained high-cadence photometry in the UV and optical to capture the early evolution. 
With these observations and the classification as a Type II supernova, we triggered our disruptive ToO program with the FUV and NUV detectors on HST/STIS.

These observations, executed 5, 6, and 7 days after explosion, show a wealth of FUV and NUV features that fade over time.
Two more observations, taken 19 and 21 days after explosion, show the (expected) complete suppression of flux in the UV.
In the first epoch, we identify narrow absorption features, many of which we attribute to both the Milky Way and host galaxy ISM.
Some narrow features are seen only at the host galaxy redshift and can either be due to CSM or ISM.
We model the supernova spectra with \texttt{CMFGEN}, finding good agreement with the $E_{\mathrm{kin}}=6\times10^{50}\mathrm{\, erg}$ explosion of a 12 \msun{} star at solar metallicity.
With this modeling we identify notable contributions from Mg, Si, C, Al, Cr, S, Ni, and Ti with the majority of the structure coming from Fe.
Using the \texttt{CMFGEN} models, we extract the photospheric temperature, radius, and velocity of SN~2022acko and compare it to SN~1999em, SN~2005cs, and SN~2006bp.
SN~2022acko falls within the range set by these supernovae for each parameter demonstrating that its photospheric properties are normal for a Type II supernova.
\citet{dessart_quantitative_2005} showed that the UV is strongly affected by metals, whose abundance is primordial at these early post-explosion epochs.
While the abundances of metals, such as Fe, can be measured through the forest of lines they produce over extended spectral regions, He, C, N, O, Mg, and Al have resonance transitions in the UV which produce strong, isolated lines which would vary with metalicity and potentially CNO processing \citep{davies_surface_2019}.
With more detailed modeling and more spectra to compare to, it is possible that these spectra could be used to measure relative metallicity (e.g. \citealt{dessart_quantitative_2005, foley_metallicity_2013}, although see also \citealt{dessart_type_2014, graham_twins_2015, derkacy_ultraviolet_2020, derkacy_sn_2022}).

We compare the spectra of SN~2022acko to the best UV data available, using spectra of SN~2005cs, SN~1999em, SN~2021yja, and SN~2022wsp. 
In these spectra we see hints of great diversity in the UV, however it is challenging to draw a true comparison given the  sparse time sampling, low S/N, host galaxy extinction, and rapid UV evolution.
Interestingly, we do not see narrow or intermediate width emission lines or P Cygni profiles associated with CSM interaction in the UV for SN~2022acko, indicating that it is either not there, at too low density to produce emission lines, or has been swept up by the ejecta already. 
Even earlier observations would further constrain the characteristics of any CSM present as well as a more detailed examination of observations at other wavelengths \citep[e.g. radio non-detection;][]{ryder_radio_2022}.

Finally, we investigate the fraction of flux observed for a variety of scenarios, finding on day 5 that 53\% of the flux is captured when the $B$-band is the bluest filter observed while 95\% of the flux is captured when observations extend to the $uvw2$ filter.
The percentage of missing flux decreases over time as the ejecta cools, highlighting the importance of even earlier UV observations, when the majority of the flux falls below 3000\AA{}.
We fit an absorbed blackbody to our full spectra, finding that the functional form of \citet{nicholl_magnetar_2017} effectively characterize the shape of the spectrum, provided the right parameters are used, matching the observed fractional flux to within 10\%.

The collection of this UV dataset took the mobilization and coordination of both ground and space-based resources with rapid response time through multiple collaborations. 
With these observations it is clear that collecting early UV spectra is worth the effort: there is a wealth of information in the UV that we are only beginning to characterize. 
Understanding the diversity of Type II supernovae in the UV will be critical for interpreting high redshift supernovae coming from the earliest stellar populations.
This can only currently be accomplished with HST. However, in the future, the rapid response of the proposed Ultraviolet Explorer \citep[UVEX][]{kulkarni_science_2021} could extend these observations even earlier, allowing us to probe the earliest phases of evolution and possible interaction with CSM.

In addition to the unique observations presented in this paper, a rich dataset has already been collected on SN~2022acko, encompassing a number of other firsts. 
It was the first supernova to use JWST observations to identify the progenitor of a supernova in HST pre-explosion images \citep[see also Kilpatrick et al., in prep.]{van_dyk_identifying_2023}  and the first core-collapse supernova for which JWST spectroscopy was obtained (see Shahbandeh et al, in prep.).
We will continue to observe SN~2022acko into the nebular phase, making it a keystone event that can be used to interpret future supernova observations.

\acknowledgments
K.A.B. thanks everyone at STScI for their work in getting this program scheduled and executed quickly, especially contact scientist TalaWanda Monroe and program coordinator Blair Porterfield.
We thank our referee for their time and comments which resulted in an improved paper.
K.A.B thanks  M. Hakan \"{O}zsara for the use of the image of NGC 1300.
This publication was made possible through the support of an LSSTC Catalyst Fellowship to K.A.B., funded through Grant 62192 from the John Templeton Foundation to LSST Corporation. The opinions expressed in this publication are those of the authors and do not necessarily reflect the views of LSSTC or the John Templeton Foundation.
Time domain research by the University of Arizona team and D.J.S.\ is supported by NSF grants AST-1821987, 1813466, 1908972, \& 2108032, and by the Heising-Simons Foundation under grant \#20201864. 
The research by Y.D., S.V., N.M. and E.H. is supported by NSF grants AST-2008108.
J.I.G.H. acknowledges financial support from the Spanish Ministry of Science and Innovation (MICINN) project PID2020-117493GB-I00.
D.J.H. thanks NASA for partial support through the astrophysical theory grant 80NSSC20K0524.
L.G. and T.E.M.B. acknowledge financial support from the Spanish Ministerio de Ciencia e Innovaci\'on (MCIN), the Agencia Estatal de Investigaci\'on (AEI) 10.13039/501100011033, the European Social Fund (ESF) ``Investing in your future'', and the European Union Next Generation EU/PRTR funds under the PID2020-115253GA-I00 HOSTFLOWS project, the 2019 Ram\'on y Cajal program RYC2019-027683-I, the 2021 Juan de la Cierva program FJC2021-047124-I, and from Centro Superior de Investigaciones Cient\'ificas (CSIC) under the PIE project 20215AT016, and the program Unidad de Excelencia Mar\'ia de Maeztu CEX2020-001058-M.
C.A. and J.D. acknowledge support by NASA grants JWST-GO-02114.032-A and JWST-GO-02122.032-A.
This work is supported by the international Gemini Observatory, a program of NSF's NOIRLab, which is managed by the Association of Universities for Research in Astronomy (AURA) under a cooperative agreement with the National Science Foundation, on behalf of the Gemini partnership of Argentina, Brazil, Canada, Chile, the Republic of Korea, and the United States of America.
E.B. was supported in part by NASA Grant 80NSSC20K0538.
The SALT data presented in this paper were obtained via Rutgers University program 2022-1-MLT-004 (PI: S.W.J.).
L.A.K. acknowledges support by NASA FINESST fellowship 80NSSC22K1599.
This paper used data obtained with the MODS spectrographs built with
funding from NSF grant AST-9987045 and the NSF Telescope System
Instrumentation Program (TSIP), with additional funds from the Ohio
Board of Regents and the Ohio State University Office of Research.
This research is based on observations made with the NASA/ESA Hubble Space Telescope obtained from the Space Telescope Science Institute, which is operated by the Association of Universities for Research in Astronomy, Inc., under NASA contract NAS 5-26555. These observations are associated with program 17132.
Support for program 17132 was provided by NASA through a grant from the Space Telescope Science Institute, which is operated by the Association of Universities for Research in Astronomy, Inc., under NASA contract NAS 5-26555.
This research was supported by the Munich Institute for Astro-, Particle and BioPhysics (MIAPbP) which is funded by the Deutsche Forschungsgemeinschaft (DFG, German Research Foundation) under Germany's Excellence Strategy - EXC-2094 - 390783311.
The INT is operated on the island of La Palma by the Isaac Newton Group of Telescopes in the Spanish Observatorio del Roque de los Muchachos of the Instituto de Astrof\'isica de Canarias.
This work makes use of observations from the Las Cumbres Observatory network.
The LCO team is supported by NSF grants AST-1911225 and AST-1911151.

J.S. acknowledges support from the Packard Foundation.

This work was based partially on observations obtained at the Southern Astrophysical Research (SOAR) telescope, which is a joint project of the Minist\'{e}rio da Ci\^{e}ncia, Tecnologia e Inova\c{c}\~{o}es (MCTI/LNA) do Brasil, the US National Science Foundation’s NOIRLab, the University of North Carolina at Chapel Hill (UNC), and Michigan State University (MSU).

This research has made use of the NASA/IPAC Extragalactic Database (NED), which is funded by the National Aeronautics and Space Administration and operated by the California Institute of Technology.
This research made use of Photutils, an Astropy package for detection and photometry of astronomical sources \citep{larry_bradley_2022_6825092}.

\vspace{5mm}
\facilities{ADS, Bok (B\&C), CTIO:PROMPT, HST (STIS), IRSA, LBT (MODS), LCOGT (SBIG, Sinistro, FLOYDS), MMT (Binospec), Meckering:PROMPT, NED, SALT (RSS), SOAR (GHTS), Swift (UVOT), TNS}

\software{ astropy \citepalias{astropy_collaboration_astropy_2013, astropy_collaboration_astropy_2018, astropy_collaboration_astropy_2022}, 
Binospec IDL \citep{kansky_binospec_2019},  BANZAI \citep{mccully_real-time_2018}, CMFGEN \citep{hillier_treatment_1998, hillier_time-dependent_2012, hillier_photometric_2019, dessart_type_2013}, FLOYDS \citep{valenti_first_2014}, HOTPANTS \citep{becker_hotpants_2015}, IDSRED \citep{tomas_e_muller_bravo_2023_7851772}, IRAF \citep{tody_iraf_1986, tody_iraf_1993}, LCOGTSNpipe \citep{Valenti2016}, Light Curve Fitting \citep{hosseinzadeh_light_2023}, MatPLOTLIB \citep{hunter_matplotlib_2007}, MESA \citep{paxton_modules_2011, paxton_modules_2013, paxton_modules_2015, paxton_modules_2018, paxton_modules_2019}, MODS pipeline \citep{pogge_rwpoggemodsccdred_2019}, NumPy \citep{harris_array_2020}, Photutils \citep{larry_bradley_2022_6825092}, PySALT \citep{crawford_pysalt_2010},  SuperBol \citep{nicholl_superbol_2018, superbol}, Scipy \citep{virtanen_scipy_2020}, V1D \citep{livne_implicit_1993}}

\appendix

\section{Photospheric Parameters for SN~2022acko}
Table \ref{tab:PhotProp} gives the temperature, radius, density, and velocity of the photosphere of the \texttt{CMFGEN} model of SN~2022acko as a function of time.
\begin{deluxetable}{lccr}
\tablecaption{Photospheric Properties of SN~2022acko\label{tab:PhotProp}}
\tablehead{\colhead{Age (d)} & \colhead{Radius ($10^{14}$ cm)} & \colhead{Velocity ($\mathrm{km\,s^{-1}}$)} & \colhead{Temperature (K)}}
\startdata
5.5 & 4.518  & 9508 & 12669\\
6.0 & 4.839  & 9335 & 11910\\
6.5 & 5.157  & 9182 & 11247\\
7.0 & 5.461  & 9029 & 10746\\
7.5 & 5.768  & 8901 & 10357\\
8.0 & 6.056  & 8761 & 10052\\
8.5 & 6.346  & 8641 & 9790\\
9.0 & 6.631  & 8529 & 9555\\
10.0& 7.185  & 8316 & 9141\\
11.0& 7.720  & 8123 & 8752\\
12.1& 8.300  & 7939 & 8349\\
13.3& 8.925  & 7761 & 7943\\
14.6& 9.569  & 7565 & 7533\\
16.1& 10.167 & 7309 & 7047\\
17.7& 10.742 & 7020 & 6603\\
19.5& 11.297 & 6712 & 6310\\
21.4& 11.835 & 6391 & 6093\\
23.6& 12.367 & 6073 & 5932\\
\enddata
\end{deluxetable}

\section{Comparison Supernovae}
\begin{deluxetable*}{lcccccr}
\tablecaption{Parameters used for Comparison Supernovae\label{tab:UVSNComp}}
\tablehead{\colhead{Supernova}   & \colhead{$E(B-V)_{MW}$} & \colhead{$E(B-V)_{host}$} & \colhead{Redshift} & \colhead{Distance} &  \colhead{Explosion Epoch} & \colhead{reference}\\
\colhead{}   & \colhead{(mag)} & \colhead{(mag)} & \colhead{} & \colhead{Modulus} &  \colhead{(JD)} & \colhead{}}
\startdata
SN~2022wsp & 0.05 & 0.3 & 0.00932 & 31.99 & \color{magenta}{2459855.08 $\pm$ 0.42}&\citealt{vasylyev_early-time_2023},\\ 
        &          &       &         &        &          & \citealt{lu_h_1993}\\ 
SN~2021yja & 0.0191  & 0.085 & 0.00568 & 31.85 & \color{magenta}{2459464.9 $\pm$0.06} & \citealt{hosseinzadeh_weak_2022}\\ 
SN~2012A   & 0.0274 & 0.009 & 0.00251 & 29.53 & \color{magenta}{2455933.5$\pm^{+1}_{-3}$} & \citealt{tomasella_comparison_2013},\\ 
        &          &       &         &        &          & \citealt{silverman_after_2017},\\ 
        &          &       &         &        &          & \citealt{tully_nearby_1988}\\
SN~2005cs  & 0.034 & 0.015 & 0.00154 &29.39  & \color{magenta}{2453549.5 $\pm$1} & \citealt{baron_reddening_2007},\\ 
        &          &       &         &        &          & \citealt{sabbi_resolved_2018},\\
        &          &       &         &        &          & \citealt{pastorello_sn_2006},\\
        &          &       &         &        &          & \citealt{silverman_after_2017}\\
SN~1999em  & 0.035 & 0.1 & 0.00239  & 30.34 & \color{magenta}{2451475.9$\pm2$} & \citealt{elmhamdi_photometry_2003},\\ 
        &          &       &         &        &          & \citealt{leonard_cepheid_2003}\\
\enddata
\end{deluxetable*}


\bibliography{SN2022acko}
\end{document}